\documentclass[pdflatex,bst/sn-mathphys-num]{sn-jnl}% Math and Physical Sciences Numbered Reference Style
\usepackage{graphicx}%
\usepackage{multirow}%
\usepackage{amsmath,amssymb,amsfonts}%
\usepackage{amsthm}%
\usepackage{mathrsfs}%
\usepackage[title]{appendix}%
\usepackage{xcolor}%
\usepackage{textcomp}%
\usepackage{manyfoot}%
\usepackage{booktabs}%
\usepackage{algorithm}%
\usepackage{algorithmicx}%
\usepackage{algpseudocode}%
\usepackage{listings}%
\usepackage{makecell}%
\usepackage{rotating}
\usepackage[utf8]{inputenc}
\usepackage{geometry}
\usepackage{multirow}
\usepackage{longtable}
\usepackage{array}
\usepackage{hhline}
\usepackage{lscape}
\usepackage{enumitem}
\usepackage{tabularray}
\usepackage{hyperref}
\usepackage{url}
\SetTblrTemplate{head}{empty}
\DefTblrTemplate{lastfoot}{default}{
  \UseTblrTemplate{caption}{default}
}

\theoremstyle{thmstyleone}%
\theoremstyle{thmstyletwo}%

\theoremstyle{thmstylethree}%

\NewDocumentEnvironment{landscapeLongTblr}{+b}
  {
    \newgeometry{
      paperwidth=\paperheight,
      paperheight=\paperwidth,
      top=2cm,
      bottom=2cm,
      left=0cm,
      right=0cm,
    }
    \begin{landscape}
    \SetTblrStyle{caption}{font = \fontsize{8pt}{9pt}\selectfont}
    \vspace*{\fill}
      #1
    \vspace*{\fill}
    \end{landscape}
    \restoregeometry
  }
  {}

\begin{document}

\title[Article Title]{A Survey on Large Language Model Impact on Software Evolvability and Maintainability: the Good, the Bad, the Ugly, and the Remedy}


\author*[1]{\fnm{Bruno} \sur{Claudino Matias}}\email{claudinomab@vcu.edu}

\author[2]{\fnm{Savio} \sur{Freire}}\email{savio.freire@ifce.edu.br}

\author[3]{\fnm{Juliana} \sur{Freitas}}\email{juliana.freitas@lsu.edu}

\author[3]{\fnm{Felipe} \sur{Fronchetti}}\email{ffronchetti@lsu.edu}

\author[1]{\fnm{Kostadin} \sur{Damevski}}\email{kdamevski@vcu.edu}

\author[1]{\fnm{Rodrigo} \sur{Spinola}}\email{spinolaro@vcu.edu}

\affil*[1]{\orgdiv{Computer Science}, \orgname{Virginia Commonwealth University}, \orgaddress{\city{Richmond}, \state{VA}, \country{United States}}}

\affil[2]{ \orgname{Federal Institute of Ceara}, \orgaddress{\state{CE}, \country{Brazil}}}

\affil[3]{\orgdiv{Computer Science and Engineering}, \orgname{Louisiana State University}, \orgaddress{\city{Baton Rouge}, \state{LA}, \country{United States}}}


\abstract{\textit{Context.}
Large Language Models (LLMs), such as GPT-4, are increasingly embedded in software engineering workflows for tasks including code generation, summarization, repair, and testing. Although empirical research shows productivity gains, better comprehension, and lower cognitive load, the evidence remains fragmented, and concerns have emerged about hallucinations, unstable outputs, methodological limitations, and 
forms of technical debt. How these mixed effects of LLM use shape long-term software maintainability and evolvability remains unclear. 

\textit{Objectives.}
This study systematically examines how LLMs influence the maintainability and evolvability of software systems. Specifically, we identify which quality attributes are addressed in existing research, the positive impacts LLMs provide, the risks and weaknesses they introduce, and the mitigation strategies proposed in the literature. 

\textit{Method.}
We conducted a systematic literature review. Searches across ACM Digital Library, IEEE Xplore, and Scopus (2020–end of 2024) resulted in 87 primary papers for in-depth analysis. Qualitative evidence was extracted through a calibrated multi-researcher process. Maintainability and evolvability attributes were examined descriptively, while positive impacts, risks, weaknesses, and mitigation strategies were synthesized using a hybrid thematic approach supported by an LLM-assisted analysis tool with human-in-the-loop validation.

\textit{Results.}
LLMs offer benefits such as enhanced analyzability, testability, code comprehension, debugging support, and automated repair effectiveness. However, they also introduce risks, including hallucinated or incorrect outputs, brittleness to context, limited domain reasoning, unstable performance, and methodological flaws in current evaluations. These structural weaknesses further threaten long-term evolvability.

\textit{Conclusion.}
LLMs can strengthen software maintainability and evolvability, but they also pose nontrivial risks that may undermine long-term sustainability. Responsible adoption requires safeguards, rigorous evaluation, and structured human oversight.}




\maketitle

\section{Introduction}\label{sec1}

Large Language Models (LLMs) such as GPT, Llama, and Claude are rapidly transforming software engineering practice, supporting activities across the development lifecycle, including code generation, repair, summarization, documentation, and analysis. Early empirical studies have demonstrated improvements in productivity and task efficiency \cite{chen2021evaluating,finnie2022copilot}, reductions in cognitive load \cite{vaithilingam2022expectation}, and enhanced comprehension and debugging support \cite{barke2023grounded,liu2023debugger}. As organizations continue integrating LLMs into their workflows, understanding their long-term effects on software quality and sustainability has become a pressing need for both researchers and practitioners.

Two core dimensions of long-term software sustainability are \emph{maintainability} and \emph{evolvability}. Maintainability refers to the ease with which software can be modified, corrected, or tested \cite{iso25010}, while evolvability concerns a system's capacity to adapt to new requirements and remain flexible over time \cite{mens2008software,ernst2022}. Although interrelated, maintainability and evolvability capture distinct aspects of software quality. Recent research suggests that LLMs may simultaneously strengthen and undermine these qualities: while LLMs can streamline documentation, reduce repetitive work, and help developers understand complex code, they may also introduce unstable changes, obscure design intent, increase hidden defects, generate hallucinated content, or add new forms of technical debt \cite{nguyen2022survey,goues2023promises,bird2024aidriven}.

%motivation

Despite the fast-growing body of research on LLM-assisted software engineering, existing evidence remains fragmented and often restricted to isolated tasks. Many empirical studies evaluate LLMs within narrow contexts, such as code generation, bug fixing, summarization, or testing, without examining how these findings relate to broader concerns about long-term software sustainability. For instance, studies on code generation frequently report productivity gains but also inconsistent or incorrect outputs \cite{chen2021evaluating,barke2023grounded}, while work on automated repair highlights both promising fixes and hallucinated patches \cite{liu2023debugger}. Research on code summarization and explanation likewise finds improvements in understandability but also notable variability and factual errors across prompts \cite{lachaux2020unsupervised,kocetkov2022stack}.

This task-specific evidence leaves unclear which maintainability or evolvability attributes are truly being encountered and addressed across the literature. Prior survey work calls attention to the scattered and preliminary nature of existing findings, noting the need for integrative perspectives on LLM impacts on software sustainability \cite{nguyen2022survey,qiao2023survey}. At the same time, researchers increasingly raise concerns about long-term risks. Recent studies warn that models can hallucinate and produce unreliable outputs~\cite{kocon2024hallucinations}, brittleness and instability of LLM-generated code \cite{barke2023grounded}, and evaluation pitfalls that complicate assessments of LLM quality in realistic settings \cite{kabir2023pitfalls}. Additional research highlights erosion of design intent, degraded traceability, and potential accumulation of new forms of technical debt when LLMs are used without adequate oversight \cite{bird2024aidriven,thaller2024cocreation}. 

Further complicating the landscape, foundational weaknesses of LLMs, including limited domain grounding, prompt sensitivity, unstable reasoning, and training-data biases, are increasingly recognized in AI and NLP research \cite{bommasani2021opportunities}, yet their implications for maintainability and evolvability are not well mapped. Although a variety of mitigation strategies have been proposed (e.g., hybrid pipelines, human-in-the-loop validation, prompt engineering, guardrails), these remain dispersed across isolated studies without systematic synthesis.

Together, these limitations make it difficult for the software engineering community to determine whether LLMs genuinely promote sustainable software development or whether their adoption risks introducing new burdens that may hinder long-term maintainability and evolvability. Addressing this gap requires a comprehensive and structured synthesis of the existing empirical evidence.

%goal

The goal of this study is to systematically investigate how LLMs influence the maintainability and evolvability of software systems, identifying the attributes they are associated with, the benefits they provide, the risks and weaknesses they introduce, and the mitigation strategies proposed in the literature.
To guide this effort, we adopt the following overarching research question:

\begin{quote}
\textit{How do Large Language Models influence the maintainability and evolvability of software systems, and what opportunities, risks, weaknesses, and mitigation strategies emerge from the current body of empirical evidence?}
\end{quote}

The overarching research question naturally aligns with the four 
perspectives used in this study: \emph{the Good, the Bad, the 
Ugly, and the Remedy}. These perspectives provide an organizing lens through which the diverse 
empirical evidence becomes interpretable. \emph{The Good} reflects opportunities 
and positive impacts identified across the literature, illustrating how LLMs can 
enhance productivity, improve code comprehension, and support sustainable 
software development practices. \emph{The Bad} represents the practical risks that 
emerge when LLMs introduce instability, incorrect outputs, or methodological 
limitations that hinder maintainability or evolvability. \emph{The Ugly} captures 
deeper structural weaknesses in current LLM technologies, such as conceptual 
misunderstanding, hallucinations, or non-generalizable behavior, that pose more 
systemic threats to long-term software sustainability. Finally, \emph{the Remedy} synthesizes the mitigation strategies proposed in the literature, focusing on how identified weaknesses can be addressed through technical, methodological, and human-centered interventions.
The specific research questions operationalizing this investigation are presented in Section~\ref{sec3}.

%method

To answer the overarching research question, we conducted a systematic literature review following established empirical software engineering guidelines~\cite{Wohlin2012, Kitchenham2015}. The review involved designing a comprehensive search strategy covering the period from 2020 to the end of 2024, applying rigorous inclusion and exclusion criteria, screening 711 studies, and ultimately selecting 87 primary papers for in-depth analysis. Evidence was analyzed using a hybrid human-in-the-loop thematic approach supported by an analysis tool, enabling the structured identification of codes and the derivation of themes from text fragments, which were then used to synthesize maintainability and evolvability attributes, positive impacts, risks, weaknesses, and mitigation strategies described in the literature.
%contributions
Therefore, this paper provides four main contributions:

\begin{enumerate}
    \item A mapping of maintainability and evolvability attributes associated with LLM-assisted software engineering.
    \item An integrated synthesis of the positive impacts LLMs may have on sustainable software development.
    \item A comprehensive identification of the risks and weaknesses that may threaten long-term maintainability and evolvability.
    \item A consolidated overview of proposed mitigation strategies that aim to address known risks and weaknesses.
\end{enumerate}

%paper structure

In addition to this introduction, this paper has six additional sections. Section~\ref{sec2} introduces the background on software maintainability and evolvability. Section~\ref{sec3} details the review methodology. Section~\ref{sec4} presents the results, and Section~\ref{sec5} provides an in-depth analysis and discussion of these findings. Section~\ref{sec6} outlines the threats to validity. Lastly, Section~\ref{sec7} concludes the paper.

\section{Background}\label{sec2}

This section provides the conceptual background necessary to contextualize our investigation. We begin by clarifying the two foundational software quality constructs, maintainability and evolvability, which underpin our analysis of how LLMs influence the sustainability of software systems over time. We then discuss the role of LLMs within the broader landscape of software engineering practice, highlighting their emerging capabilities, limitations, and relevance to activities that shape long-term system quality. Together, these elements establish the conceptual grounding for the subsequent review and synthesis.

\subsection{Maintainability and Evolvability}
Clarifying the conceptual foundations of \textit{software maintainability} and \textit{software evolvability} is essential for this study, as these constructs underpin how we understand the long-term sustainability of software systems. Although often used interchangeably, the two terms reflect distinct yet complementary perspectives. 

As Buckley et al.~\cite{Buckley2005} observe, ``unlike evolvability, the term maintainability is closely associated with the term legacy system,'' highlighting the importance of treating them separately.
Maintainability generally refers to post-delivery activities aimed at preserving a system's correctness and operational quality, such as bug fixing, performance tuning, and minor enhancements. Evolvability, in contrast, concerns the system's ability to accommodate substantive and continuous change throughout its life cycle. It captures the ease with which new features, architectural adjustments, or design improvements can be introduced while maintaining long-term adaptability.

Rajlich~\cite{Rajlich2014} further expands this distinction by framing evolvability as a broader and more forward-looking concept than traditional maintenance. While much of the evolvability literature emphasizes structural and architectural properties, it also implicitly encompasses human and process factors that shape how systems evolve over time. From this perspective, maintainability is a short-term quality attribute associated primarily with corrective and adaptive modifications, whereas evolvability represents a longer-term view focused on a system's sustained capacity for change, growth, and architectural adaptability.

Breivold et al.~\cite{Breivold2010} describe evolvability as a composite quality property comprising analyzability, integrity, changeability, extensibility, portability, and testability. Earlier frameworks and standards, such as ISO/IEC 9126~\cite{ISO9126}, often conflate maintainability and evolvability by grouping them into compound attributes such as modularity, reusability, analyzability, modifiability, and testability.

As summarized in Table~\ref{tab:attributes}, the attributes used to classify maintainability and evolvability quotes share several conceptual intersections, but they also exhibit meaningful distinctions. Stability and maintainability compliance are \textit{exclusive} to maintainability, reflecting concerns with preventing unintended side effects during modifications and ensuring adherence to established standards. In contrast, integrity and extensibility appear \textit{only} under evolvability, emphasizing architectural coherence and the capacity to introduce new features with minimal disruption.

Some attributes, such as changeability, testability, and portability, have identical formal definitions in both constructs. As discussed in Section~\ref{sec4}, their practical classification depends heavily on the context in which the attribute is invoked within each study. Analyzability presents the most notable divergence: within maintainability, it focuses on diagnosing failures or identifying components that require modification, whereas in evolvability it concerns identifying which parts of the system will be affected by current or future change stimuli. These nuances illustrate how similar attributes can play different roles depending on whether the analytical lens emphasizes short-term maintenance activities or long-term system evolution.

\begin{table}[t]
    \centering
    \begin{tabular}{|p{0.2\textwidth} | p{0.35\textwidth} | p{0.35\textwidth}|} \Xhline{3\arrayrulewidth}
        \textbf{Attribute} & \textbf{Maintainability Definition} & \textbf{Evolvability Definition} \\ \Xhline{3\arrayrulewidth}
        Analyzability & The capability of the software product to be diagnosed for deficiencies or causes of failures in the software, or for the parts to be modified to be identified.etc.~\cite{ISO91261abc} & The capability of the software system to enable the identification of influenced parts due to change stimuli. The change stimuli include changes in business model, business objectives, functional and quality requirements, environment, underlying technologies and emerging technologies, new standards, new infrastructure, etc.~\cite{evolvabilityAttributes} \\ \hline
        Changeability & The capability of the software product to enable a specified modification to be implemented.~\cite{ISO91261abc} & The capability of the software system to enable a specified modification to be implemented. ~\cite{evolvabilityAttributes}\\ \hline
        Stability & The capability of the software product to avoid unexpected effects from modifications of the software.~\cite{ISO91261abc} & N/A \\ \hline
        Testability & The capability of the software product to enable modified software to be validated.~\cite{ISO91261abc} & The capability of the software product to enable modified software to be validated.~\cite{evolvabilityAttributes} \\ \hline
        Maintainability compliance & The capability of the software product to adhere to standards or conventions relating to maintainability.~\cite{ISO91261abc} & N/A \\ \hline
        Portability & The capability of the software product to be transferred from one environment to another.~\cite{ISO91261abc} & The capability of the software system to be transferred from one environment to another.~\cite{evolvabilityAttributes}  \\ \hline
        Integrity & N/A & The capability of the software system to maintain architectural coherence while accommodating changes.~\cite{evolvabilityAttributes} \\ \hline
        Extensibility & N/A  & The capability of the software system to enable the implementation of extensions to expand or enhance the system with new capabilities and features with minimal impact to existing system. Extensibility is a system design principle where the implementation takes into consideration of future growth.~\cite{evolvabilityAttributes}\\ \Xhline{3\arrayrulewidth}
    \end{tabular}
    \caption{Comparison of maintainability and evolvability attributes.}
    \label{tab:attributes}
\end{table}

\subsection{Large Language Models in Software Engineering}

Large Language Models have rapidly become influential tools in software engineering, enabling natural language interactions across a wide range of development activities. Models such as GPT-3, GPT-3.5, and GPT-4, LLaMA~2, CodeLlama, StarCoder, and CodeT5~\cite{wang2021codet5} have demonstrated strong capabilities in code generation, repair, summarization, explanation, and reasoning, reshaping how developers engage with software artifacts. Empirical studies show that LLM-based assistants can support developer productivity and code comprehension. For example, Nair et al.~\cite{nair2023copilot} report that GitHub Copilot helps professional developers complete tasks more efficiently and with reduced cognitive load. Similarly, Sarkar et al.~\cite{sarkar2023llmsummarization} find that LLM-based approaches substantially improve code summarization accuracy, supporting program understanding tasks.

Beyond productivity and comprehension, LLMs have been applied to testing, debugging, and program repair. Xiang et al.~\cite{xiang2023testgen} demonstrate that LLM-generated test cases can achieve competitive coverage and fault-detection performance. Work on automated program repair (APR) shows that LLMs can generate correct patches for a subset of defects and outperform certain traditional APR techniques~\cite{xia2022repair, sobania2023largerepair}. Research has also extended LLM usage to requirements engineering, where Zhang et al.~\cite{zhang2023llmre} show that LLMs can assist in requirements classification and traceability with promising levels of accuracy.

Despite these advances, important limitations remain. Studies document hallucinations and incorrect code generation~\cite{chen2021codex, vaithilingam2022copilotstudy}, sensitivity to prompts and context~\cite{wei2022chainofthought}, training-data contamination and leakage~\cite{liu2023trainingdata}, and challenges related to correctness and security~\cite{pearce2022copilotsecurity}. Broader analyses of LLM-based SE research highlight reproducibility issues, insufficient reporting, and methodological inconsistencies~\cite{nguyen2023reproducibility, barke2023copilotresearch}. These weaknesses raise concerns regarding the reliability, robustness, and long-term implications of LLM-generated artifacts.

Overall, while LLMs are increasingly adopted across diverse SE tasks, the empirical landscape remains fragmented. Existing studies tend to focus on isolated activities, such as code generation, summarization, repair, or testing, without examining how these findings relate to broader concerns about software sustainability. In particular, there is no consolidated understanding of how LLMs influence the long-term maintainability and evolvability of software systems, despite these being foundational quality attributes that shape a system’s capacity to remain adaptable, reliable, and cost-effective over time. This gap underscores the need for a comprehensive synthesis that integrates existing evidence and clarifies the implications of LLM adoption for sustainable software engineering.

\section{Methodology}\label{sec3}

This study follows established guidelines for the transparent reporting of empirical research in software engineering. In particular, we adopt established practices for structuring and presenting the research protocol as recommended by Wohlin et al.~\cite{Wohlin2012} and Kitchenham et al.~\cite{Kitchenham2015}. These practices ensure methodological rigor, replicability, and clarity in describing our objectives, design, data collection, and analysis procedures. Accordingly, this section details the research questions, search strategy, data extraction and analysis procedures.

\subsection{Research Questions}

%The rapid integration of LLMs into software engineering workflows raises fundamental questions about their broader implications for long-term software quality. 
To structure our investigation into how LLMs influence key sustainability attributes, particularly maintainability and evolvability, we define a set of focused research questions. These questions provide a coherent lens through which to interpret the diverse findings reported in the selected studies.

\begin{itemize}
    \item \textbf{RQ1:} What maintainability and evolvability attributes are addressed by the use of LLMs in software engineering tasks?
    \item \textbf{RQ2 (The Good):} What positive impacts do LLMs provide for software evolvability and maintainability?
    \item \textbf{RQ3 (The Bad):} What weaknesses do LLMs exhibit with respect to software evolvability and maintainability?
    \item \textbf{RQ4 (The Ugly):} What weaknesses exist within LLMs concerning software evolvability and maintainability?
    \item{\textbf{RQ5 (The Remedy):} How can weaknesses of LLMs be mitigated to
better support software evolvability and maintainability?}
\end{itemize}

These research questions reflect the need to holistically assess the benefits, risks, and drawbacks of adopting LLMs across software development activities. \textbf{RQ1} focuses on identifying which maintainability and evolvability attributes are addressed when LLMs are used in software engineering tasks, providing an evidence-based understanding of how different quality attributes surface in the literature. \textbf{RQ2} examines the enabling role of LLMs in improving software maintainability and evolvability throughout the software life cycle. \textbf{RQ3} investigates the possible risks, such as technical debt accumulation, dependency issues, or loss of traceability, that may threaten long-term maintainability and evolvability. \textbf{RQ4} concentrates on the weaknesses of current LLMs, such as hallucinations, limited domain understanding, and challenges in code correctness, that can hinder their effective use in evolving software systems. Finally, \textbf{RQ5} explores the strategies proposed in the literature to mitigate these weaknesses. Together, these research questions provide a structured lens through which to analyze the interplay between LLM technologies and the sustainability of evolving software systems.

\subsection{Search Strategy}

The first step of our literature review consisted of identifying and selecting the relevant studies. To ensure comprehensive coverage and alignment with the research questions, we defined a structured full-text search strategy based on key concepts central to our investigation: (i) the use of large language models (LLMs), (ii) their application context in software systems, and (iii) their relationship with software evolvability and maintainability. The search was manually executed using the following search string:

\begin{verbatim}
(large language model OR LLM OR ChatGPT OR LLaMa OR CoPilot)
AND (software OR system OR application)
AND (maintainability OR maintenance OR evolvability OR evolution)
\end{verbatim}

The first group of terms, ``large language model,'' ``LLM,'' and the specific names of representative tools (``ChatGPT,'' ``LLaMa,'' ``CoPilot''), captures both the general concept and the most widely adopted models, ensuring inclusion of papers referring to LLMs generically or by name. The second group, ``software,'' ``system,'' and ``application'', delimits the technological and engineering domain of interest, excluding unrelated uses of LLMs (e.g., in linguistics or education). The final group, ``maintainability,'' ``maintenance,'' ``evolvability,'' and ``evolution'', narrows the scope of the search to studies that explicitly address the core constructs of maintainability and evolvability, which underpin this review. This combination of keywords thus balances recall and precision, maximizing the likelihood of retrieving studies relevant to the multifaceted impact of LLMs on software evolvability and maintainability.

At this stage, we also selected three major digital libraries as our primary data sources: \textit{ACM Digital Library}, \textit{IEEE Xplore}, and \textit{Scopus}. These venues were chosen because they jointly provide broad, high-quality coverage of software engineering research \cite{Petersen2015}. ACM DL and IEEE Xplore index the flagship conferences and journals of the software engineering community, ensuring access to authoritative and peer-reviewed empirical studies. Scopus, in turn, aggregates publications from a wide range of publishers and offers comprehensive indexing of interdisciplinary work. Using this combination of databases helps make our search more comprehensive and consistent with recommended practices for secondary studies in software engineering.

\subsection{Inclusion and Exclusion Criteria}

To ensure that the selected papers were directly relevant to our research objectives, we defined a set of inclusion and exclusion criteria before conducting the screening process. Our goal was to identify studies that report on the use or evaluation of LLM-based tools, techniques, or products that address aspects of software evolvability or maintainability, and that are situated within the broader context of the software development life cycle (SDLC). The criteria were established to capture contributions with empirical or conceptual relevance to the topic while filtering out works lacking methodological grounding or relevance to software engineering. The criteria are presented below:

\begin{itemize}[label={\tiny$\bullet$}, leftmargin=*]
    \item \textbf{Inclusion Criteria 1 (IC1):} Paper investigates software evolvability or maintainability in relation to the use of LLMs;
    \item \textbf{Inclusion Criteria 2 (IC2):} The study presented in the paper must be related to software development activities (e.g., design, implementation, testing, documentation, etc.);
    \item \textbf{Inclusion Criteria 3 (IC3):} Paper is available in English;
    \item \textbf{Exclusion Criteria 1 (EC1):} Paper is a duplicate or a conference paper extended into a journal article; 
    \item \textbf{Exclusion Criteria 2 (EC2):} Paper is not in the software engineering domain;
    \item \textbf{Exclusion Criteria 3 (EC3):} Paper is available for download;
    \item \textbf{Exclusion Criteria 4 (EC4):} Abstracts, presentations.
\end{itemize}

All inclusion and exclusion criteria were applied systematically to the entire set of retrieved papers. One researcher conducted the initial screening by evaluating each paper against all criteria, ensuring consistent application throughout the corpus. A second researcher independently reviewed these decisions to verify correctness and identify potential misclassifications. Any disagreements or ambiguities were discussed collaboratively until consensus was reached, ensuring that only studies meeting the predefined methodological requirements were included in the final set. This two-stage screening and reconciliation process strengthened the reliability and transparency of the study selection phase.

\subsection{Data Extraction}

After finalizing the set of primary studies, we conducted a systematic data extraction process to ensure reliability, consistency, and transparency. The main objective of this phase was to collect evidence from each paper that could be compared and synthesized in relation to our research questions. To minimize potential threats to validity, such as individual interpretation bias, inconsistent coding, and data omission, the extraction procedure followed five steps:

\begin{enumerate}
    \item \textbf{Division of workload:} The complete corpus of selected papers was divided into four approximately equal subsets, each assigned to a different researcher. This distribution reduced individual workload and allowed parallel yet independent data collection, mitigating risks of fatigue-related bias and ensuring diverse analytical perspectives.
    
    \item \textbf{Calibration and alignment:} Prior to the full extraction, two iterative calibration rounds were conducted on two representative papers. During these sessions, all researchers jointly extracted and discussed the same data points to harmonize interpretation, resolve ambiguities, and ensure a shared understanding of the constructs of maintainability, evolvability, and LLM-related impacts. This calibration step was critical to establish a common coding frame and reduce construct interpretation bias.
    
    \item \textbf{Independent extraction:} Each of the four researchers independently extracted data from their assigned subset using a standardized extraction template. Independence at this stage ensured that interpretations were not influenced by group dynamics or dominant voices, thereby strengthening the internal validity of the collected data.
    
    \item \textbf{Cross-review and adjudication:} One researcher was designated as adjudicator and systematically reviewed all extracted data across subsets. This cross-validation ensured terminological consistency, confirmed adherence to the predefined criteria, and resolved any discrepancies through comparison and clarification. The adjudication process acted as an additional control mechanism to ensure uniformity and coherence across the entire dataset.
    
    \item \textbf{Resolution of disagreements:} Any remaining doubts or conflicting interpretations were resolved through one-on-one discussions between the researchers involved. When consensus could not be reached, decisions were revisited collectively during team meetings. These dialogues served both to refine the coding schema and to ensure that all decisions were traceable and explicitly documented.
\end{enumerate}

\noindent
For each paper, we collected the following data items:

\begin{itemize}[label={\tiny$\bullet$}, leftmargin=*]
    \item \textbf{General information:} Title, publication year, and application domain;
    \item \textbf{LLM-related information:} Name of the LLM or model family, and the specific software life cycle activity addressed (accompanied by verbatim quotations to preserve contextual meaning);
    \item \textbf{Maintainability attributes} (based on ISO/IEC 9126~\cite{ISO9126}): Analyzability, Changeability, Stability, Testability, Maintainability Compliance, and Portability;
    \item \textbf{Evolvability attributes} (based on Breivold et al.~\cite{Breivold2010}): Analyzability, Integrity, Changeability, Portability, Extensibility, and Testability;
    \item \textbf{Impact dimensions:} 
    \begin{itemize}
        \item \textit{(The Good)} Positive impacts and advantages of LLMs, with supporting quotations;
        \item \textit{(The Bad)} Potential risks emerging from LLM adoption, with quotations;
        \item \textit{(The Ugly)} Identified weaknesses of LLMs, with quotations.;
        \item \textit{(The Remedy)} Proposed mitigation strategies, with quotations.
    \end{itemize}
\end{itemize}

To further strengthen dependability and confirmability, all extracted data were stored in a shared spreadsheet accessible to all team members. Each entry was accompanied by the exact text excerpt from the source article to ensure traceability and enable later verification during synthesis. The combination of independent extraction, adjudication, and consensus-building discussions minimized subjectivity and enhanced the overall validity and reproducibility of this phase.

%It is important to note that a single paper could contain multiple relevant quotations. For maintainability and evolvability, each quotation could also map to more than one attribute when its content aligned with multiple definitions. The categories \textit{Positive Impacts}, \textit{Potential Risks}, and \textit{Weaknesses} were analyzed using thematic synthesis. %Finally, some identified weaknesses did not have corresponding solutions, as several studies reported problems without proposing any mitigation strategies.

\subsection{Data Analysis}

Our data extraction produced two types of data. The first type comprises descriptive data, such as LLM names and software development activities. These elements are straightforward to categorize. We also treated maintainability and evolvability attributes as descriptive data, since we applied predefined labels and definitions during extraction.

The second type consists of qualitative data, specifically the quotes extracted for \textit{Positive Impacts}, \textit{Risks}, \textit{Weaknesses}, and \textit{Mitigation Solutions for Weaknesses}. These excerpts provide richer textual evidence and require more careful interpretation, as different quotes may express similar ideas using different wording. To analyze this data, we conducted a thematic analysis guided by methodological recommendations described in \cite{seaman_qualitative}, which defines standards for synthesizing qualitative evidence in software engineering. In addition, we explored the use of an LLM as a supplementary aid to support the early organization of codes during synthesis. The LLM did not replace researchers’ judgments. Instead, it generated candidate codes and themes from the extracted quotations within a human-in-the-loop process, where researchers validated outputs and provided textual feedback to refine subsequent generations. This exploratory use reflects current limitations and open methodological questions regarding the appropriateness and reliability of LLMs for qualitative analysis in empirical software engineering~\cite{11071418}.

\subsubsection{Rationale for Integrating LLMs into the Coding and Thematic Analysis}

The coding and thematic analysis phase requires synthesizing a diverse set of qualitative excerpts. Conducting this process manually introduces substantial challenges related to consistency, scalability, and cognitive load. Qualitative coding demands that the researcher maintain a stable interpretive lens across all excerpts while iteratively refining emerging themes. This makes it difficult to ensure both \textit{local coherence} (accurate representation of individual excerpts) and \textit{global coherence} (consistent coding and terminology across the entire corpus). When performed solely by humans, this process often results in overlapping or redundant codes, inconsistent vocabulary, and repeated cycles of refinement that become increasingly inefficient as the dataset grows \cite{miles2014qualitative, saldana2021coding, braun2006using}.

To mitigate these challenges, we incorporated LLM as a support tool for generating initial codes and candidate themes. Recent work has shown that LLMs can assist qualitative analysis by identifying semantic patterns across heterogeneous text and promoting terminological consistency during early-stage coding ~\cite{11071418}. These models help reduce the mechanical effort involved in comparing conceptually similar excerpts expressed in different ways, thereby lowering the likelihood of redundant codes and minimizing the need for multiple rounds of reorganization.

Importantly, the LLM was used as an assistive mechanism rather than an autonomous decision-maker. All suggested codes, themes, and higher-level categories were inspected, refined, or merged by the research team to ensure methodological rigor and theoretical alignment. This hybrid approach, leveraging the LLM’s ability to surface semantic regularities while preserving human interpretive oversight, provided assistance in organizing the analysis in a more coherent manner, while we remained attentive to preserving the qualitative foundations of the study.

Lastly, to promote transparency and methodological rigor in our use of LLMs, we adhered to the guidelines proposed in \cite{wagner2025evaluationguidelinesempiricalstudies} for integrating LLMs into empirical research. These guidelines emphasize the importance of clearly declaring the use of an LLM and specifying its role in the study, identifying the exact model and version employed, and detailing the configuration parameters used during analysis. The authors also recommend reporting the prompts adopted in the procedure, prioritizing the use of open LLM models whenever possible, and ensuring that all LLM-generated outputs receive thorough human review and validation.

\subsubsection{LLM-ThemeCrafter Tool}

Building upon these guidelines, as well as established procedures for thematic synthesis in software engineering \cite{Cruzes2011recomendedstepsforthematicsynthesisinse}, we developed a tool that systematically feeds the extracted quotes into an LLM and replicates the key steps of the thematic synthesis process. The tool generates candidate codes, themes, and higher-level themes for each collection of quotes. The complete implementation and all associated resources are available in our GitHub repository \cite{ThemeCrafter2025}.

\begin{figure}[!h]
    \centering
    \includegraphics[width=1\linewidth]{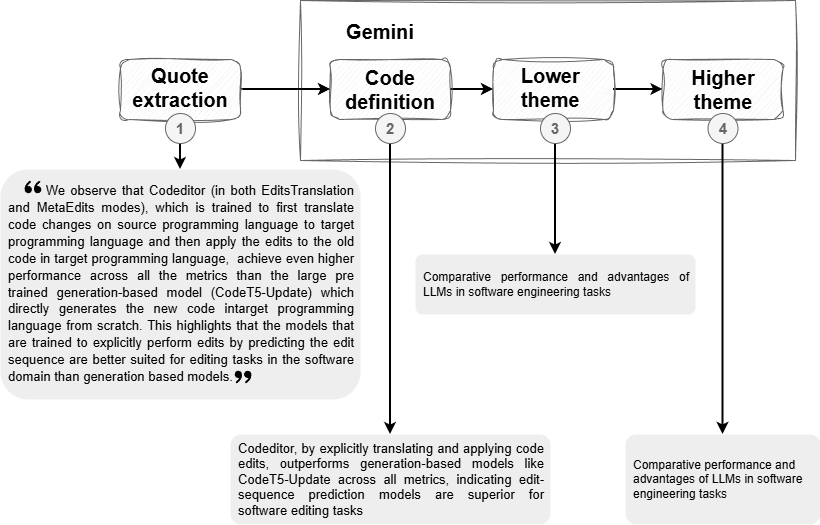}
    \caption{Example illustrating the synthesis pipeline from manual quote extraction to LLM-generated codes and themes made by LLM-ThemeCrafter.}

    %\caption{Diagram showing the flow of synthesis in one of the extracted quotes. The quote is analyzed by the LLM that generate a code, a quicker description trying to capture its main information, with a human-in-the-loop approach so every result can validated. Then all the codes generated are passed a second time through the LLM so it can generate around 15 lower themes that groups all the quotes. Lastly, the process is repeated so we can end with around 5 high order themes that encapsulates all the original quotes.}
    \label{fig:quote_diagram_positive_impact}
\end{figure}

\noindent\textbf{Step 1 - Quote extraction.}
This step was completed during the data extraction phase, in which we collected quotations from the selected papers. For the thematic synthesis, the quotations were organized in a spreadsheet and grouped by subject, allowing the LLM to read them and perform the subsequent steps appropriately.

%The tool -- called LLM Thematic Synthesizer --.

\vspace{\baselineskip}

\noindent\textbf{Step 2 - Code Generation.}
The first step performed by the tool is reading the configuration file, which initializes all environment settings. Afterward, the tool loads the spreadsheet containing the extracted quotes and invokes the LLM client responsible for generating the initial codes.  The method that handles code generation also accepts two additional parameters: \verb|humanInTheLoop| (default = \verb|False|) and \verb|chunkSize| (default = 10), both specified in the \textit{config.yaml} file. The \textit{chunkSize} parameter defines how many quotes are processed by the LLM at once, and the tool iteratively calls the Gemini model until all chunks have been coded. This mechanism helps manage the number of tokens sent to the LLM, an important consideration depending on the model being used, and integrates with our human-in-the-loop approach. When \textit{humanInTheLoop} is set to \verb|True|, each generated chunk of codes is printed to the terminal, and the user is asked to provide feedback by responding with one of the following options:

\begin{itemize}
    \item \textbf{[Y]es}: Confirms that the generated chunk is approved.
    \item \textbf{[A]ll}: Approves the current chunk and all subsequent chunks. Used when the LLM is consistently producing satisfactory codes.
    \item \textbf{[N]o}: Rejects the generated chunk. The researcher is then prompted to provide feedback.
\end{itemize}

When a chunk is rejected, the user provides textual feedback, after which the tool regenerates the chunk incorporating the supplied guidance. This feedback, along with the original rejected chunk and a timestamp, is stored in a \textit{feedback.csv} file. These records can later be used to inform future code generation and to support visual inspection and validation.

% \begin{lstlisting}[language=Python, caption={Prompt used for code quotes.}, label={lst:prompt_code}]
% """You are a content analyzer. You will receive an array with multiple quotes from papers on Software engineering.
% Each position is a cell of a spreadsheet, and each cell represents one quote from some paper.

% For each position of the array, return a JSON object where the key is the array position (0, 1, 2, ...),
% and the value is a list of codes. Each code should follow the Thematic Synthesis Code step proposed in
% 'Recommended Steps for Thematic Synthesis in Software Engineering'.

% <example output omitted>

% Here are some quotes and codes for examples:

% <few-shot examples >

% Always return a JSON with the same number of keys as the input array.
% If you cannot generate a code for one position, return an empty array [] for that position.
% If you think there is one quote that should not be there or gives too little information, put N/A as a text. Ex: "11": ["N/A"]
% Be careful not to switch codes.
% Return ONLY valid JSON. Do not add explanations.
% Here are the quotes:"""
% \end{lstlisting}

Finally, we incorporate an additional layer of contextual information into the prompt. In this context-injection step, we use Python’s \textit{PdfReader} library to extract the relevant sections from the paper \textit{Recommended Steps for Thematic Synthesis in Software Engineering} \cite{Cruzes2011recomendedstepsforthematicsynthesisinse}. These excerpts provide the LLM with explicit methodological guidance, ensuring that its coding and theme-generation process aligns with established best practices for conducting thematic synthesis in software engineering.

\vspace{\baselineskip}

\noindent\textbf{Steps 3 and 4 - Themes and Higher-level Themes Generation.}
The generated codes are stored as a new column in the spreadsheet, after which the tool proceeds with a similar process to create themes and higher-level themes. Unlike the code-generation stage, this step does not employ the human-in-the-loop mechanism, because we did not observe issues with the LLM grouping the codes it had previously produced. However, theme generation is still processed in fixed-size chunks. Early experiments showed that creating themes required substantially more computational resources, which caused timeouts when using less capable models such as \verb|gemini-2.5-flash|. Introducing a chunking strategy allowed the tool to perform theme generation reliably while still using the free \verb|gemini-2.5-flash| model. The tool also maintains a global data structure that records each generated theme and the number of codes associated with it. This information is later added to the spreadsheet as a separate column and is used as input for generating higher-level themes.

% To summarize how the synthesis pipeline operates from raw evidence to conceptual themes, Figure~\ref{fig:quote_diagram_positive_impact} presents a worked example using one quote extracted from~\ref{}. The diagram depicts a four-step flow:

% \begin{enumerate}
%     \item \textbf{Quote Extraction (Manual)}: A relevant excerpt is manually identified and extracted from the primary study by the researchers. In the example, the selected quote describes how Codeditor, a model that translates and applies code edits, achieves superior performance compared to generation-based models when supporting software editing tasks.

%     \item \textbf{Code Definition}: Using the LLM, the quote is transformed into a concise analytic code that captures its core meaning. Here, the model condenses the original passage into a focused statement emphasizing Codeditor's strengths and its comparative advantages over CodeT5-Update.

%     \item \textbf{Lower-Level Theme Generation}: The analytic code is grouped with similar codes to form an initial, low-level theme. In this case, the resulting theme highlights the \textit{comparative performance and advantages of LLMs in software engineering tasks}.

%     \item \textbf{Higher-Level Theme Construction}: Finally, the lower-level theme is abstracted and merged into a broader, higher-order theme that represents a recurrent pattern across multiple studies. For this example, the higher-level theme remains centered on the comparative advantages of LLMs, illustrating how themes accumulate into generalizable insights.
% \end{enumerate}

To show how the synthesis pipeline progresses from raw evidence to conceptual themes, Figure~\ref{fig:quote_diagram_positive_impact} presents a worked example based on a single quote extracted from ~\cite{A60}. The process begins with manual quote extraction, where the researchers identify a relevant passage from a primary study. In the example shown, the selected quote explains how Codeditor, a model designed to translate and apply code edits, outperforms generation-based models when supporting software editing tasks. The extracted quote is then converted into a concise analytic code using the LLM. This step distills the original passage into a focused statement that captures its essential meaning, emphasizing Codeditor’s strengths and its comparative advantages over CodeT5-Update. Next, the analytic code is grouped with semantically similar codes to form a lower-level theme. In this case, the resulting theme characterizes the comparative performance and advantages of LLMs in software engineering tasks. Finally, the lower-level theme is further abstracted and integrated into a higher-level theme that reflects a recurring pattern across multiple studies. In this example, the higher-level theme continues to emphasize the comparative advantages of LLMs, illustrating how individual observations accumulate into more general insights.

% Through this example, the figure demonstrates how raw textual evidence is progressively structured, from a detailed quote, to an analytic code, and then to increasingly abstracted themes, ultimately supporting the synthesis of broader insights about the positive impacts of LLMs in software engineering activities.

The tool, named LLM-ThemeCrafter, was implemented in Python, a language well suited for rapid prototyping and supported by a rich ecosystem of libraries for LLM interaction and data manipulation. We selected Gemini (model \verb|gemini-2.5-flash|) as the LLM for this work due to its free availability. The tool currently expects the spreadsheet containing the extracted quotes to be stored in Google Sheets. While this introduces a dependency, Google Sheets is freely accessible and its configuration for read–write permissions is simple.

\section{Results}\label{sec4}

\begin{figure}[t]
    \centering
    \includegraphics[width=1\linewidth]{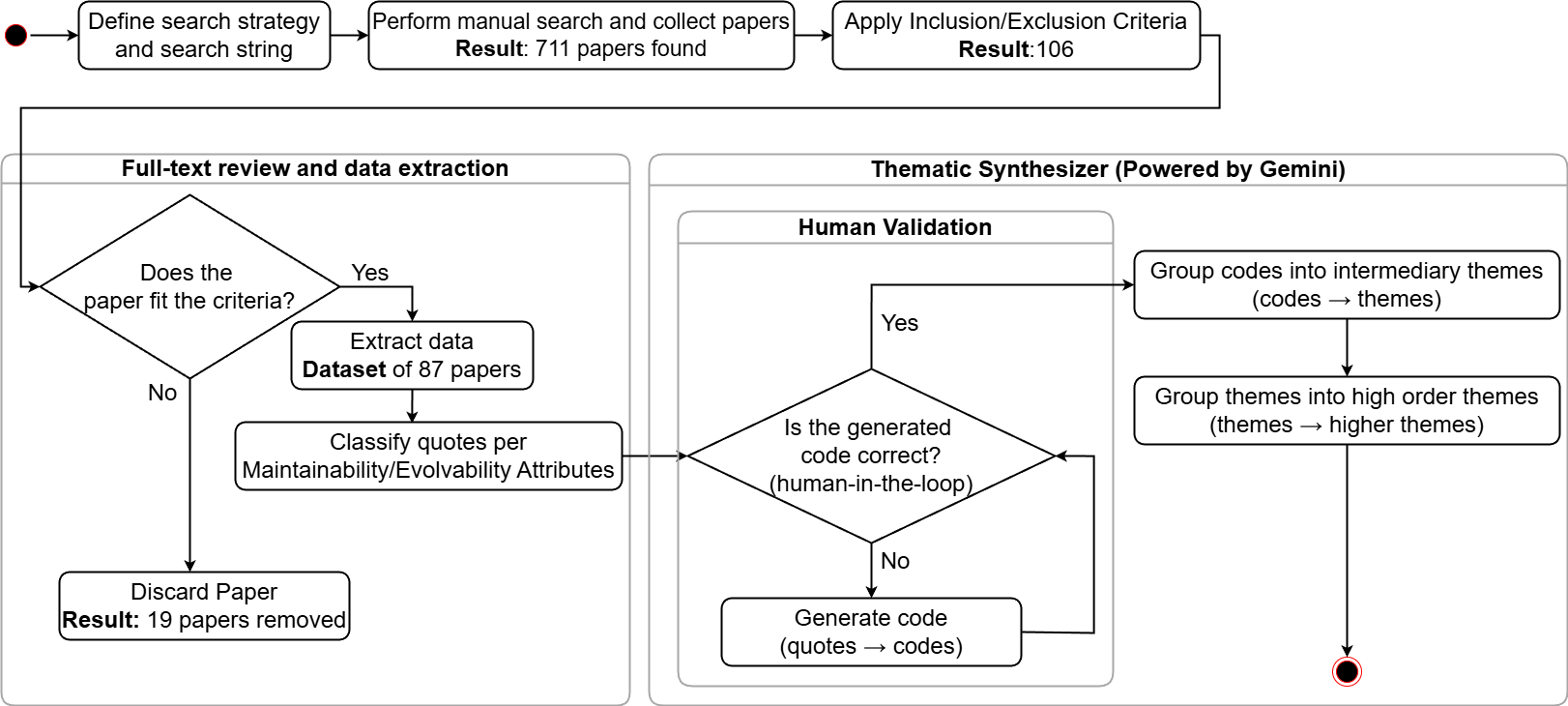}
    \caption{Followed systematic literature review workflow.}
    \label{fig:workflow}
\end{figure}

The overall executed procedure is illustrated in Figure~\ref{fig:workflow}. The workflow is organized into two major phases: (1) identifying and selecting relevant studies, and (2) extracting, coding, and synthesizing qualitative evidence. We began by defining the search strategy. After querying the selected digital libraries, we retrieved an initial set of 711 papers. Each paper was screened using the predefined inclusion and exclusion criteria, and studies that did not meet these criteria, or were identified as duplicates, were removed. This initial screening excluded 605 papers, leaving 106 studies for full-text review and data extraction.

During the full-text reading phase, an additional 19 papers were excluded because they did not provide sufficient evidence for extraction, did not actually report empirical results relevant to our research focus, or upon closer examination did not align with the scope defined in our inclusion criteria. This screening resulted in a final set of 87 papers.

During the data extraction phase, we collected relevant quotes and metadata from each included study and used them as input for the LLM-ThemeCrafter tool. The tool generated preliminary codes, which were reviewed and refined by a researcher using a human-in-the-loop validation strategy to ensure fidelity to the original evidence. Validated codes were then processed by the LLM-ThemeCrafter to produce intermediary themes that grouped related concepts, which were subsequently synthesized into higher-order themes for broader abstraction. 

\subsection{Demographics}

The primary studies included in this review were collected from a combination of major digital libraries and key software engineering conferences. Using the search string defined in Section \ref{sec3}, we retrieved 24 papers from the \textit{ACM Digital Library}, 37 from \textit{IEEE Digital Library}, and 26 from \textit{Scopus}. 

The review literature covered studies published up to the end of 2024. However, no relevant papers addressing the relationship between LLMs and software maintainability or evolvability were identified prior to 2023. Most studies were published in 2024 with 72\% of the total number of papers, showing a rapid expansion of research and widespread adoption of large language models in software engineering.

To contextualize the landscape of research included in our review, we first provide a descriptive overview of the analyzed studies. This demographic analysis characterizes the papers along several dimensions relevant to understanding how LLMs are being applied within software evolution/maintenance. In particular, we examine the distribution of LLMs used and the software development life cycle (SDLC) activities targeted by the proposed approaches. 

\subsubsection{LLM Usage Landscape}

This subsection analyzes which large language models were employed in the selected studies. When a study reported the use of multiple LLMs, each model was counted separately. In the same way, studies that mentioned the LLM family, but did not specify the model, were marked as ``Unspecified". Across the dataset, a total of 123 LLMs usage instances were identified. The aggregated distribution of LLM usage is presented in Table~\ref{tab:llmUse}.

\begin{table}[ht]
    \centering
    \begin{tabular}{|c | c | c|}
        \Xhline{3\arrayrulewidth}
        \textbf{LLM Family} & \textbf{Model} & \textbf{Usage Count}\\ \Xhline{3\arrayrulewidth}
        \multirow{10}{0.1\textwidth}{GPT} & GPT 3.5 Turbo & 17 \\ \hhline{~--}
         & Unspecified & 14 \\ \hhline{~--}
         & GPT 3.5 & 12 \\ \hhline{~--}
         & GPT 4 & 11\\ \hhline{~--}
         & GPT 3 & 3\\ \hhline{~--}
         & GPT-J & 3\\ \hhline{~--}
         & GPT-Neo & 2\\ \hhline{~--}
         & GPT 2 & 1\\ \hhline{~--}
         & GPT 4 Turbo & 1\\ \hhline{~--}
         & GPT-NeoX & 1\\ \hline
         \multirow{6}{0.1\textwidth}{LLaMA} & CodeLLama & 4 \\ \hhline{~--}
         & LLaMA-2 & 4\\ \hhline{~--}
         & llama2-7b & 2\\ \hhline{~--}
         & KG-llama2-7b & 2\\ \hhline{~--}
         & LLaMA 3 & 1\\ \hhline{~--}
         & Unspecified & 1\\ \hline
         \multirow{4}{0.1\textwidth}{BERT} & CodeBERT & 3 \\ \hhline{~--}
         & GraphCodeBert & 3\\ \hhline{~--}
         & BERT & 1\\ \hhline{~--}
         & Sentence-BERT & 1\\ \hline
         \multirow{3}{0.1\textwidth}{T5} & CodeT5 & 4 \\ \hhline{~--}
         & T5 & 2\\ \hhline{~--}
         & CoditT5 & 1\\ \hline
         & Copilot & 6\\ \hline
         & Codex & 5\\ \hline
         & Claude & 2\\ \hline
         & CodeGen & 2\\ \hline
         & CodeParrot & 1\\ \hline
        DeepSeek & DeepSeek-Coder 6.7B & 1\\ \hline
         & CodeUp & 1\\ \hline
         & INCODER & 1\\ \hline
         & Magicoder & 1\\ \hline
         & UniXcoder & 1\\ \hline
         & Galpaca & 1\\ \hline
         & Zephyr & 1\\ \hline
         & Google PaLM & 1\\ \hline
         & Mistral-7B-Instruct & 1\\ \hline
         & Mixtral & 1\\ \hline
         & StarChat-\(\beta\) & 1\\ \hline
         & Replit-Code & 1\\ \hline
         & Custom LLM & 1\\ \hline
    \end{tabular}
    \caption{LLM usage landscape.}
    \label{tab:llmUse}
\end{table}

The results show a strong concentration around a small number of models, specially OpenAI's ChatGPT family, which is by far the most frequently used, accounting for 65 of the 123 reported instances (52.85\%). Among these, GPT-3.5 Turbo is the most referenced variant, followed by studies that do not specify the exact model version. This dominance likely reflects ChatGPT's accessibility and widespread popularity not only in academia, but also in the industry.

The second most frequently used family is LLaMA, with 14 instances, primarily involving LLaMA-2 and CodeLLaMA. Bert-based models appear in eight studies, mainly through CodeBERT and GraphCodeBERT. Other models including T5-based architectures, Github, Copilot and Codex are reported less frequently.

All remaining LLM families appeared only sporadically. Notably, none of the selected studies report the use of Google Gemini/Bard models. Although Gemini (then branded as Bard) was officially released in 2023, API access for developers only became available in December of that year. Because our study selection includes papers published up to the end of 2024, the short window between the model’s release and the cutoff date likely limited opportunities for researchers to incorporate Gemini into their studies, which may explain its absence from the dataset.

At the same time that it identifies the dominant technologies, these results also expose notable gaps in reporting practices. Many studies do not specify the exact model or version used, limiting transparency and complicating reproducibility. The absence of Gemini/Bard models further reflects the constraints imposed by release timelines and early API access rather than a lack of relevance. Taken together, these findings suggest that LLM-based SE research is still shaped by a small set of dominant, well-supported models, and that clearer reporting guidelines and broader experimentation with diverse LLM families could improve the robustness and comparability of future studies.

\subsubsection{Software Development Life Cycle Activity}\label{sec:SDLC}

Understanding which stages of the SDLC are represented in the selected studies clarifies the types of software engineering contexts and activities from which the evidence about maintainability and evolvability is drawn. Our search strategy specifically targets research that links LLMs to these quality attributes, the SDLC activities in the corpus delineate the task environments and usage scenarios to which the findings for RQ1–RQ5 apply. To characterize this context, we categorized all reported activities using the SDLC taxonomy proposed in~\cite{wang2025softwaredevelopmentlifecycle}, which organizes activities into six main groups and several subgroups. Using this scheme, we mapped the 81 activities described across the analyzed papers to their corresponding categories. The results are shown in Table~\ref{tab:SLDC}

\begin{longtblr}[
    caption={Table showing all software development life cycle activity landscape.},
    label={tab:SLDC}
]{
    colspec = {Q[l,m,4cm] X[l,t] Q[c,m,2cm] Q[c,m,1cm]},
    vlines,
    hlines,
    hline{1-2}= {2pt,solid}
}
\textbf{Activity}  & \textbf{Sub-Activity} & \textbf{Number of Papers} & \textbf{Total} \\ 

\SetCell[r=6]{} Software Development & Code Summarization & 9 & \SetCell[r=6]{} 28\\
& Code Quality Refinement & 7 &\\
& Code Generation & 4 &\\ 
& Code Understanding and reasoning & 4 &\\ 
& Code Competition & 3 &\\ 
& Commit Message Generation & 1 &\\ 

\SetCell[r=8]{} Software Maintenance & Program Repair & 13 & \SetCell[r=8]{} 28\\
& Bug Localization & 4 &\\
& Log Parsing/Analysis & 3 &\\
& Code Search & 2 &\\
& Vulnerability Repair & 2 &\\
& Fault Diagnosis & 2 &\\
& Log Generation & 1 &\\
& Documentation / Knowledge Management & 1 &\\

\SetCell[r=4]{} Testing & Test Generation & 9 & \SetCell[r=4]{} 15\\
& Vulnerability Detection & 4 &\\
& Validation & 1 &\\
& Defect Detection & 1 &\\

\SetCell[r=3]{} Requirement Engineering & Classification & 5 & \SetCell[r=3]{} 8\\
& Traceability & 2 &\\
& Elicitation & 1 &\\

Cross-stage & CI/CD Pipelines Integration & 2 & 2

\end{longtblr}

Overall, the SDLC analysis shows that the studies included in our review predominantly apply LLMs to code-centric development and maintenance activities, with comparatively fewer investigations focused on testing, requirements engineering, or multi-phase workflows. This distribution does not characterize LLM research in software engineering broadly, but it clarifies the kinds of tasks represented in our targeted corpus linking LLMs to maintainability and evolvability. Software development and software maintenance were the most represented activities, each accounting for 28 out of the 81 classified activities (34.57\%). Within software development, the largest subgroup was code summarization (9 papers), followed by code quality refinement (7 papers). Code generation and code understanding and reasoning each appeared in 4 papers, while code completion (3 papers) and commit message generation (1 paper) were less frequently explored.

For software maintenance, program repair was the most frequent subgroup, appearing in 13 papers. Bug localization (4 papers) and log parsing/analysis (3 papers) followed, while code search, vulnerability repair, and fault diagnosis each appeared in 2 papers. Log generation and documentation/knowledge management were the least represented subgroups, each appearing once.
Testing was the third most common SDLC category, with 15 papers (18.52\%). Most of these studies focused on test generation (9 papers), with others examining vulnerability detection (4 papers), validation (1 paper), and defect detection (1 paper).

Requirements engineering accounted for 8 papers (9.88\%), primarily centered on classification tasks (5 papers), with smaller attention to traceability (2 papers) and elicitation (1 paper).
Finally, cross-stage activities, those spanning multiple phases of the SDLC, were the least frequent, appearing in only 2 papers (2.47\%). Both studies addressed CI/CD pipeline integration, reflecting early efforts to apply LLMs in end-to-end or multi-phase development workflows.

\subsection{RQ1 - What maintainability and evolvability attributes are addressed by the use of LLMs in software engineering tasks?}

To answer RQ1, we analyzed the maintainability and evolvability attributes reflected in the empirical evidence reported by the selected studies. The extracted quotes are unevenly distributed across attributes, revealing clear patterns in how current LLM-based approaches emphasize particular aspects of software quality. 

\subsubsection{Maintainability Attributes}

In total, we mapped 68 quotations related to maintainability attributes. As shown in Table~\ref{tab:maintainability_attributes}, analyzability accounts for 29 quotes (42.67\%), representing nearly half of all maintainability-related observations. This emphasize LLMs as tools for supporting program understanding, diagnosis and inspection of software and related artifacts. The most frequent scenarios will include code summarization, explanation of logic or assistance in identifying defects as supported by the findings in section~\ref{sec:SDLC}.

Testability is shown as the second most frequently addressed attribute, with 12 quotes. These studies highlights the effort of using LLMs in test cases, improve fault detection or support validation activities, suggesting that LLMs are increasingly being explored in testing-related maintenance tasks. Closely following is maintainability compliance, with 11 quotes, reflecting explicit effort to align LLM-generated artifacts with coding standards, guidelines or best practices.

Stability appears in 8 quotes and relates to concerns about preserve system behavior after LLM assisted changes, avoiding unintended side effects. Compared with analyzability and testability, stability had more limited evidences, indicating fewer studies explicitly evaluating the long time robustness of LLM-generated modifications. Changeability is less represented, with only 5 quotes, suggesting relative small attention is paid to how easily those systems can be modified beyond isolated tasks. The least addressed maintainability attribute, appearing on only 3 quotes, is portability, suggesting a very low interest in LLM-based research for cross-environment or cross platform systems.

Overall, the maintainability attributes points out towards short-term activities usually related to code and other software artifacts, while attributes related to long-term modifications, robustness and environmental flexibility received limited attention if compared with the others.

\begin{table}[!ht]
    \centering
    \begin{tabular}{|c|c|} \Xhline{3\arrayrulewidth}
        \textbf{Attribute} & \textbf{Number of quotes} \\ \Xhline{3\arrayrulewidth}
        Analyzability & 29 \\ \hline
        Testability & 12 \\ \hline
        Maintainability compliance & 11 \\ \hline
        Stability & 8 \\ \hline
        Changeability & 5 \\ \hline
        Portability & 3 \\ \hline
    \end{tabular}
    \caption{Distribution of maintainability attributes identified across the analyzed studies}
    \label{tab:maintainability_attributes}
\end{table}

\subsubsection{Evolvability Attributes}
Evolvability evidence is less abundant with a total of 45 quotes, and the distribution across attributes is different when compared with maintainability. As shown in Table~\ref{tab:evolvability_attributes}, changeability is the most frequently observed evolvability attribute with 12 quotes (25\%). This suggest that when studies do address evolvability they tend to focus on how LLMs support modifications and adaptations of existing systems, such as generation patches, evolving functionalities or adjusting code.

Analyzability is a close second, with 11 quotes (22.9\%), indicating that understanding which parts of a system are impacted by changes is also a recurring concern. Integrity accounts for 10 quotes (20.8\%), reflecting attention to architectural coherence and preservation of consistency when changes are introduced with LLM assistance. This highlights an awareness on LLM-generated changes may affect the whole system structure.

Extensibility appears in 9 quotes (18.8\%), showing some level of concern in using LLMs to allow future features or expanding functionalities. In contrast, testability and portability are minimally represented with only 3 quotes each (6.3\%), suggesting that these attributes are rarely shown concern explicitly in evolvability in the literature.

\begin{table}[!ht]
    \centering
    \begin{tabular}{|c|c|} \Xhline{3\arrayrulewidth}
        \textbf{Attribute} & \textbf{Number of quotes} \\ \Xhline{3\arrayrulewidth}
        Changeability & 12 \\ \hline
        Analyzability & 11 \\ \hline
        Integrity & 10 \\ \hline
        Extensibility & 9 \\ \hline
        Testability & 3 \\ \hline
        Portability & 3 \\ \hline
    \end{tabular}
    \caption{Distribution of evolvability attributes identified across the analyzed studies}
    \label{tab:evolvability_attributes}
\end{table}

\subsection{RQ2 - What positive impacts do LLMs provide for software evolvability and maintainability?}

To address RQ2, we analyzed quotation from primary studies where authors explicitly describe the benefits or positive outcomes achieved by their proposed LLM-based approaches. In total, 140 quotes were identified and synthesized through thematic analysis, resulting in five higher-level themes for positive impacts. Table~\ref{tab:positive_impacts} presents an overview of those themes, exemplary quotes that fit the respective high-level theme and the ID of papers in which we found quotes for that theme.

The five higher-level themes are presented below. The distribution of quotes across these themes suggests that the literature emphasizes both versatility of LLMs across tasks and their concrete performance when applied to code-centric activities:

\begin{enumerate}
    \item Broad Application and Adaptability of LLMs in Software Engineering: 43 quotes
    \item Factors, Optimization, and Behavioral Aspects of LLMs: 37 quotes
    \item LLM Performance in Code Generation, Analysis, and Repair: 35 quotes
    \item Impact on Developers and Software Development Lifecycle: 18 quotes
    \item Evaluation and Testing of AI/Automated Systems in Software Engineering: 7 quotes
\end{enumerate}

The most prominent theme (\textit{Broad Application and Adaptability of LLMs in Software Engineering}) highlights that authors point out that \textbf{LLMs can be applied to a wide range of activities, often being compared and outperforming traditional techniques or baselines models}. The analyzed studies frequently rely on standard evaluation metrics (e.g., BLEU, METEOR, ROUGE-L) or direct comparisons with alternative LLMs and non-LLM approaches, reporting improvements in accuracy, efficiency, or task completion rates.

\textbf{Several studies emphasize the effectiveness of LLMs in natural language processing tasks relevant to software engineering, such as summarization, classification, and mapping between natural language and code artifacts}. Additional evidence highlights the ability of LLMs to operate in zero-shot or few-shot settings, allowing them to support tasks even in the absence of labeled data. Although less frequent, some studies also report promising  results in highly structured or model-driven evolution scenarios, suggesting that LLMs may extend beyond purely text-based tasks.

Collectively, this theme underscores LLMs as flexible tools that can be integrated into diverse stages of the software lifecycle, very frequently achieving positive results, particularly in contexts where understanding transforming or reasoning about software artifacts is required.

The theme  \textit{Factors, Optimization, and Behavioral Aspects of LLMs} focuses on how authors analyze, refine and optimize their LLM behavior to achieve better outcomes for their specific approaches. Most evidence in this category relates to strategies for improving LLM performance, such as prompt engineering, parameter tuning, and the use of complementary techniques to guide or constrain model outputs. Beyond optimization techniques, several studies also report observations about LLM behavior, including prompt formulation, trade-offs between performance and output stability, and variability across different tasks settings. This theme suggest that \textbf{positive impacts depend on careful configuration and adaptation of the models specific for the problem being addressed}. From a maintainability and evolvability perspective, this highlights that it is required effort to achieve LLM benefits and outcomes that are consistent and reliable over time.

The third theme (\textit{LLM Performance in Code Generation, Analysis, and Repair}) captures positive impacts that directly concern software artifacts generated or modified by LLMs. Evidence in this category emphasizes improvements in code quality, correctness and repair effectiveness. Several studies report that LLMs can generate higher quality code, identify and fix defects efficiently, or outperform traditional automated program repair techniques under certain conditions.

This category include improvements in identifying and fixing coding issues, robustness and reliability of LLM-assisted solutions, and comparative advantages in automated program repair. Although less frequently reported, some studies also highlighted gains in code coverage and security-related improvements. Together these findings suggest that \textbf{LLMs can positively influence both corrective and adaptive maintenance tasks, potentially reducing developer effort and improving the health of evolving code bases}.

Another theme is \textit{Impact on Developers and Software Development Lifecycle} which focuses on the impact of LLMs on developer and development processes. Findings in this category highlights \textbf{improvements in productivity, reductions on workload, faster tasks completion and lower operational costs when combined with the use of AI}. \textbf{LLMs are often described as assistants that handle routine or repetitive tasks, allowing developer to focus on higher level reasoning and design activities}.

%These benefits may be relevant from the maintainability and evolvability perspective, as they suggest that LLMs indirectly contribute to software sustainability by reducing cognitive load improving developer efficiency. However, these impacts are most often described in terms of developer productivity and workflow efficiency, without explicitly linking them to concrete maintainability or evolvability attributes of the software artifacts.

Although \textit{Evaluation and Testing of AI/Automated Systems in Software Engineering} is supported by fewer quotes, this theme includes evidence that LLMs can assist in testing LLM-based system themselves or in evaluating automated software engineering tools. Studies in this category report effective use of automated testing techniques, fault detection, and validation mechanisms, suggesting that \textbf{LLMs can play a role not only as development aid, but also as components within broader quality}.

Overall, the literature describe LLMs as powerful and versatile tools. In particular, a substantial portion of the evidence emphasizes improvements in code comprehension, debugging, and repair, as captured by the themes related to LLM performance in code generation, analysis, and repair, as well as their broad applicability across software engineering tasks. Additionally, several studies report productivity gains and workflow improvements, which are discussed under the theme addressing impacts on developers and the software development life cycle. At the same time, the results consistently show that these benefits depend on careful configuration, optimization strategies, and contextual adaptation of LLMs, as highlighted by the evidence grouped under factors, optimization, and behavioral aspects. Taken together, these findings suggest that while \textbf{LLMs demonstrate substantial potential to support sustainable software development, their positive impacts are unevenly distributed across contexts and tasks, and cannot be assumed without appropriate safeguards, motivating the examination of risks and limitations in subsequent research questions}.

Table~\ref{tab:theme_attribute_pi} synthesizes how the positive-impact themes identified in RQ2 map onto the maintainability and evolvability attributes adopted in this study. The mapping shows that the themes broad application and adaptability of LLMs, LLM performance in code generation, analysis, and repair, factors, optimization, and behavioral aspects of LLMs, and evaluation and testing of AI/automated systems in software engineering are strongly associated with \emph{analyzability}, \emph{testability}, and \emph{maintainability compliance} within the maintainability construct. This reflects a recurring emphasis on program comprehension, validation activities, and adherence to standards aimed at preserving structural coherence during maintenance tasks. For the evolvability construct, \emph{changeability} and \emph{extensibility} emerge as the most prominent attributes, indicating a prevailing concern with supporting future system growth and adaptation. In contrast, the theme impact on developers and the software development lifecycle exhibits a distinct pattern for maintainability, being primarily associated with \emph{stability}, which underscores how productivity gains and workflow support are frequently discussed in terms of preserving system behavior while improving understanding.

\begin{landscapeLongTblr}
\begin{longtblr}[
    caption={Table showing all higher themes for the quotes on positive impacts and a few examples of quotes that would fit their themes. The `ID' column shows the id for the paper from where the quote was extracted. The full table of selected studies and their references is available in \cite{selectedStudies}.  },
    label={tab:positive_impacts}
]{
    colspec = {Q[l,m,2cm] Q[l,t,2cm] X[j,m]},
    vlines,
    hlines,
    hline{1-2}= {2pt,solid},
    rows = {font = \fontsize{8pt}{9pt}\selectfont}
}
\textbf{Higher-level Theme}  & \textbf{ID} & \textbf{Exemplary Quotes} \\ 
% -- First High Order Theme
\SetCell[r=3]{} Broad Application and Adaptability of LLMs in Software Engineering
& \SetCell[r=3]{} A8, A24, A26, A28, A29, A30, A34, A35, A40, A44, A48, A50, A54, A59, A60, A62, A63, A76, A77, A79, A80, A81, A85, A88, A90, A92, A95, A96, A97, A98, A99
& ``This suggests that GPT-3.5 can be used to classify issue reports without any task-specific fine-tuning with a negligible loss in performance compared to BERT-like models, which is a significant advantage in the absence of labeled data"[A54]
\\ 

&
& ``the LLM received a high volume and variety of requests from different users and domains, indicating that the LLM was popular and useful for various language-related tasks"[A24]
\\ 

&
& ``our solution is easier to use, maintain, and deploy in IDEs."[A35]
\\ 

% -- Second High Order Theme
\SetCell[r=3]{} Factors, Optimization, and Behavioral Aspects of LLMs
& \SetCell[r=3]{} A4, A6, A8, A11, A16, A23, A34, A36, A41, A42, A53, A54, A59, A60, A63, A66, A68, A70, A74, A77, A81, A83, A88, A93, A94, A105, A106
& ``We observe that if we apply the model longer and generate more samples, we can drastically improve the number of correct bugs fixed in all three datasets and achieve very close result to that obtained by the best baseline."[A16]
\\ 

& 
& ``Furthermore, we observe that keeping the label explanation is useful to yield better performance for the 1-2-shot settings. However, we also observe the models give more non-sense responses in such settings."[A54] \\ 

& 
& ``AI-driven refactoring shows promise, particularly for smaller projects, addressing challenges related to project size, non-compilable code, and resource constraints is essential for broader applicability"[A4] \\

% -- Third High Order Theme
\SetCell[r=3]{} LLM Performance in Code Generation, Analysis, and Repair
& \SetCell[r=3]{} A5, A8, A9, A12, A15, A16, A23, A31, A34, A35, A43, A45, A51, A58, A61, A62, A68, A69, A70, A73, A74, A77, A78, A79, A83, A89, A90, A100
& ``The research utilizing ChatGPT for bug fixing and vulnerability detection presents several advantages, such as the potential to identify and rectify coding issues efficiently."[A9]
\\ 

& 
& ``In comparison to the Codex system, which achieved a bugfixing rate of 21 out of 40 bugs, ChatGPT surpassed it by successfully repairing 25 out of 40 bugs in the QuixBugs benchmark without any specific bug information, and further improved its performance by fixing 30 out of 40 bugs with detailed bug descriptions. Additionally, when compared to standard automatic program repair techniques, ChatGPT outperformed them significantly, fixing 7 out of 40 bugs. These findings highlight the potential of ChatGPT as a powerful tool for enhancing code quality and reducing the burden of manual bug fixing. The ability of ChatGPT to understand and generate human-like code explanations contributes to its effectiveness in identifying and resolving complex programming errors."[A9]
\\ 

& 
& ``Our results indicate that, with the improvements in newer versions of Copilot, the percentage of vulnerable code suggestions has reduced from 36.54\% to 27.25\%."[A12]
\\
% -- Fourth High Order Theme
\SetCell[r=3]{} Impact on Developers and Software Development Lifecycle
& \SetCell[r=3]{} A3, A15, A31, A35, A40, A47, A51, A62, A74, A83, A85, A86, A87, A91, A101, A102
& ``Improves software understading and offers levels of abstraction; better documentation and user comprehension without relying on developer's comments."[A3]
\\ 

& 
& ``developers preferred our OMG-generated commit messages as the final commit messages for 61.4\% of the code changes, while they preferred FIRA only for 10.5\% and human-written for 28.1\%. Overall, OMG generates commit messages with higher Rationality, Comprehensiveness, and Expressiveness than the FIRA and human written message."[A62]
\\ 

& 
& ``Support for repetitive and routine tasks through “quick commands”. Quick commands are shortcuts offered by StackSpot AI, which developers could use to automate common software engineering tasks, such as creating tests, documenting code, or even asking the tool to explain a certain code snippet."[A83]
\\ 

% -- Fifth High Order Theme
\SetCell[r=3]{} Evaluation and Testing of AI/Automated Systems in Software Engineering
& \SetCell[r=3]{} A23, A46, A61, A84, A103, A104
& ``CCTEST identifies numerous defects when being used to test (commercial) code completion systems, despite the varying thresholds used in deciding outliers."[A23]
\\ 

& 
& ``State-of-the-art paraphrasing techniques can be used as starting point to test the robustness of DL based code recommenders, since they are able to generate semantically equivalent descriptions of a reference text in up to 77\% of cases."[A61]
\\ 

& 
& ``Our evaluation demonstrates that FIXCHECK is capable of generating fault-revealing tests, showcasing the incorrectness of 62\% of patches authored by developers."[A46]
\\ 
\end{longtblr}
\end{landscapeLongTblr}

\begin{longtblr}[
    caption={Table mapping positive impacts themes to maintainability and evolvability quality attributes. Numbers in parentheses represent the number of primary studies within each theme associated with a given attribute.},
    label={tab:theme_attribute_pi}
]{
    colspec = {Q[l,m,6cm] Q[l,t,2cm] X[l,m]},
    vlines,
    hlines,
    hline{1-2}= {2pt,solid},
    rows = {font = \fontsize{8pt}{9pt}\selectfont}
}
\textbf{RQ2 Theme}  & \textbf{Quality Construct} & \textbf{Related Attributes} \\ 

% -- First High Order Theme
\SetCell[r=2]{} LLM Performance in Code Generation, Analysis, and Repair
& Maintainability
& Analyzability (14), Maintainability compliance (9), Testability(6), Stability (3), Changeability (2), Portability (2) \\

& Evolvability
& Extensibility (5), Changeability (4), Analyzability (2), Portability (2), Integrity (2), Testability(1) \\

% -- Second High Order Theme
\SetCell[r=2]{} Factors, Optimization, and Behavioral Aspects of LLMs
& Maintainability
& Analyzability (12), Maintainability compliance (6), Testability(5), Changeability (4), Portability (3), Stability (2) \\

& Evolvability
& Changeability (6), Analyzability (6), Integrity (6), Extensibility (3), Portability (2), Testability(1) \\

% -- Third High Order Theme
\SetCell[r=2]{} Broad Application and Adaptability of LLMs in Software
& Maintainability
& Analyzability (17), Testability(7), Maintainability compliance (5), Changeability (4), Portability (2), Stability (2) \\

& Evolvability
& Changeability (10), Extensibility (6), Integrity (6), Analyzability (4), Portability (2), Testability(1) \\

% -- Fourth High Order Theme
\SetCell[r=2]{} Impact on Developers and Software Development Lifecycle
& Maintainability
& Stability (8), Analyzability (7), Testability(4), Maintainability compliance (4), Changeability (2), Portability (1)  \\

& Evolvability
& Changeability (5), Extensibility (4), Integrity (3), Analyzability (1), Portability (1) \\

% -- Fith High Order Theme
\SetCell[r=2]{} IEvaluation and Testing of AI/Automated Systems in Software Engineering
& Maintainability
& Analyzability (3), Testability(2), Maintainability compliance (2), Changeability (2), Portability (1)  \\

& Evolvability
& Changeability (2), Extensibility (1), Integrity (1) \\

\end{longtblr}

\subsection{RQ3 -- What potential risks emerge from the utilization of LLMs for software evolvability and maintainability?}

To answer RQ3, we analyzed quotations that explicitly report negative impacts, risks or limitations associated with the use of LLMs in software engineering tasks, resulting in 48 quotes extracted. The higher-level themes are presented below, and in Table~\ref{tab:risks} we present exemplary quotes and papers related to this findings.

\begin{enumerate}
    \item Methodological and Evaluation Challenges in LLM Research: 14 quotes
    \item LLM Output Quality, Accuracy, and Maintainability: 14 quotes
    \item Practical Deployment and Contextual Adaptation Challenges: 14 quotes
    \item Ethical, Security, and Trustworthiness Concerns in LLM Usage: 3 quotes
    \item Impact of Training Data on AI Model Performance: 3 quotes
\end{enumerate}

Overall, the identified risks reveal that limitations associated with LLM adoption are not confined to isolated tasks or domains. Instead, they span methodological practices, artifact quality, deployment contexts, and data-related concerns, collectively posing challenges to long-term software maintainability and evolvability.

A prominent portion of the reported risks concerns \textit{methodological and evaluation challenges in LLM research}, which account for 29.16\% of the total quotations. Several studies highlight that current evaluation practices may inadequately capture the true behavior of LLM-based approaches. Some examples include the use of datasets that are outdated, insufficiently diverse, or potentially included in model training raises concerns about data leakage and threats to external validity. Another methodological weakness identified is the reliance on automated metrics without complementary qualitative validation, which can obscure defects in LLM-generated artifacts and complicate the interpretation of reported performance gains. These findings suggest that \textbf{methodological weaknesses may lead to overly optimistic conclusions regarding the sustainability of LLM-assisted solutions, or poorly constructed solutions when the provided context is not clearly presented to the LLM}.

The second major theme relates to \textit{LLM output quality, accuracy, and maintainability}, emphasizing risks directly affecting the software artifacts produced with LLM assistance. Multiple studies report incorrect, misleading, or hallucinated outputs, including instances where a majority of generated solutions were found to be wrong. Beyond correctness, concerns also emerge regarding the maintainability of LLM-generated code and documentation, which may be unnecessarily complex, inconsistent, or difficult for developers to understand and evolve over time. Such behaviors \textbf{introduce instability into evolving systems, increases debugging and correction effort, and may accumulate new forms of technical debt}.

\textit{Practical deployment and contextual adaptation challenges} form another substantial category of risk. Evidence indicates that \textbf{LLM performance is highly sensitive to context quality, domain knowledge, and prompt construction}. This theme includes performance problems in the output regarding how hard was the problem, and if the context provided was adequate to find a correct solution, i.e. providing good essential and specific context to the LLM, not overwhelming it on information. Moreover, it highlights that the improvement on productivity was achieved in scenarios where the developer knew how to use correctly the AI tool. These factors can undermine the long-term viability of LLM-based workflows, particularly in large or safety-critical systems.

A smaller set of risks arises from the \textit{impact of training data on model behavior}. Studies report that low-quality, biased, or unrepresentative training or fine-tuning data can propagate errors and reinforce undesirable patterns in generated artifacts. Over time, such issues \textbf{may negatively affect both maintainability and evolvability by embedding flawed assumptions or unstable behaviors into software systems}.

Finally, \textit{ethical, security, and trustworthiness concerns} highlight risks that extend beyond technical correctness. These include the inadvertent exposure of sensitive information, generation of insecure code, and amplification of biases present in training data or code repositories. Although less frequently reported, these risks \textbf{have potentially severe consequences, reinforcing the need for safeguards and human oversight when incorporating LLM outputs into production systems}.

Taken together, the findings indicate that \textbf{while LLMs offer potential benefits, their use also introduces a diverse set of risks that may threaten long-term software maintainability and evolvability. These risks underscore the importance of robust evaluation practices, careful deployment strategies, and governance mechanisms to prevent the accumulation of technical and ethical debt in evolving software systems}.

Table \ref{tab:theme_attribute_risks} synthesizes how the risk-related themes identified in RQ3 map onto the maintainability and evolvability attributes considered in this study. The mapping indicates that themes associated with unreliable or incorrect LLM outputs, limited robustness, and workflow-related risks are predominantly linked to \emph{stability}, \emph{analyzability}, and \emph{maintainability compliance} within the maintainability construct. This highlights recurring concerns about unintended side effects of LLM-assisted changes, increased difficulty in diagnosing faults, and potential misalignment with established development standards. From an evolvability perspective, the identified risk themes are mainly associated with \emph{changeability}, \emph{integrity}, and \emph{extensibility}, reflecting apprehensions that LLM usage may hinder controlled system evolution, erode architectural coherence, or introduce changes that complicate future adaptations. Overall, the table clarifies that the risks reported in RQ3 extend beyond isolated task failures and manifest as threats to core maintainability and evolvability attributes, emphasizing the potential long-term impact of unmanaged LLM adoption on software sustainability.

\begin{landscapeLongTblr}
\begin{longtblr}[
    caption={Table showing all higher themes for the quotes on potential risks and a few examples of quotes that would fit their themes. The `ID' column shows the id for the paper from where the quote was extracted. The full table of selected studies and their references is available in \cite{selectedStudies}.},
    label={tab:risks},
]{
    colspec = {Q[l,m,2cm] Q[l,t,2cm] X[j,m]},
    vlines,
    hlines,
    hline{1-2} = {2pt,solid},
    rows = {font = \fontsize{8pt}{9pt}\selectfont}
}
\textbf{Higher-level Theme}  & \textbf{ID} & \textbf{Exemplary Quotes} \\ 
    
% -- First High Order Theme
\SetCell[r=3]{} Methodological and Evaluation Challenges in LLM Research
& \SetCell[r=3]{} A4, A28, A29, A42, A61, A62, A63, A70, A87, A95, A103, A105
& ``Additionally, software projects that have no clear identifiable components or modules make it difficult to determine how the segments should be chosen."[A4]
\\ 

& 
& ``Validation with a single dataset poses an external validity threat. Classification performance may differ in other commit datasets."[A28]
\\ 

& 
& ``The training corpus of ChatGPT includes open-source projects before Sep. 2021. Thus there may be data leakage, i.e., ChatGPT may have seen the commit messages for the test cases during its pre-training."[A29]
\\

% -- Second High Order Theme
\SetCell[r=3]{} LLM Output Quality, Accuracy, and Maintainability
& \SetCell[r=3]{} A45, A46, A47, A70, A77, A86, A88, A89, A90, A91, A93, A102, A104
& ``Finding 7: ChatGPT-generated code contains various types of code style and maintainability
issues. Their common issues are specific to the language and tool being used."[A70]
\\ 

& 
& ``Despite the very successful applications of ChatGPT to process-related tasks, we observed cases of what has been recently defined as artificial hallucination [73], namely confident responses provided by an AI such as ChatGPT which look plausible to the human interacting with it but that are clearly wrong."[A77] \\ 

& 
& ``However, we also found that ChatGPT struggles on tasks involving refining documentation and functionalities, mainly due to a lack of domain knowledge, unclear location, and unclear changes in the review comments."[A40] \\

% -- Third High Order Theme
\SetCell[r=3]{} Practical Deployment and Contextual Adaptation Challenges
& \SetCell[r=3]{} A44, A54, A70, A83, A94, A96, A97, A98, A99, A100, A101, A106
& ``Task difficulty, time that tasks are introduced, and program size impact automated
code generation performance: We found that the performance of ChatGPT on code generation
tasks is significantly influenced by factors such as task difficulty, task-established time, and program size. This suggests that improvements in AI models should consider these factors to better
adapt to different types of code generation tasks."[A70]
\\ 

& 
& ``Productivity gains unlock only if users know how to use
StackSpot AI. A participant commented that, like any AI tool,
StackSpot AI only brings time-saving benefits if used correctly with
refined prompts and proper settings. If used incorrectly, it could
even lead to time wastage.
"[A83]
\\ 

& 
& ``Figuring out what is a good knowledge source. As mentioned in
Section 3.1, knowledge sources are representative documents that
enrich the prompts for RAG’s generation component, providing
essential context for task development. Without these sources, responses from StackSpot AI would be less contextualized, resembling
the answers from general-purpose coding AI assistants. Thus, identifying effective knowledge sources is vital for StackSpot AI’s performance. "[A83]
\\ 

% -- Fourth High Order Theme
\SetCell[r=3]{} Impact of Training Data on AI Model Performance
& \SetCell[r=3]{} A48, A80
& ``The quality of human-written commit messages may
also affect the performance of models for commit message generation. "[A80]
\\ 

& 
& ``Most existing commit message generation
models, especially the ones using machine learning techniques,
were trained on datasets consisting of human-written commit
messages. If these datasets include poor-quality messages, the
models might learn to replicate these inadequacies. This is a
critical concern, as the models might generate commit messages
that are technically similar to the training data but are not useful
in practice. For the task of commit message generation, there
is still a lack of a high-quality dataset. Therefore, existing
models for commit message generation have certain quality
risks, which expect more attention from the researchers."[A80]
\\ 

& 
& ``Without in-domain training data, the F1-score achieved by ChatGPT for anomaly detection is only 0.450, making it unsuitable for direct application in industry."[A48]
\\ 

% -- Fifth High Order Theme
\SetCell[r=3]{} Ethical, Security, and Trustworthiness Concerns in LLM Usage
& \SetCell[r=3]{} A12, A76, A77
& ``The issues associated with the security of generated code, especially from LLMs, will continue to impact the quality of code generation tools and thus might reduce the trust of developers using such tools. It is important to continue investigating such issues as both the underlying code generation models and the nature of weaknesses evolve fast."[A12]
\\ 

&
& ``UniLog may encounter ethical issues associated with LLM usage [7], namely (1) Using third-party LLM models as a backbone may result in privacy leakage on codes or logs. (2) Utilizing codes containing stereotypes or racial biases as candidates may induce generating harmful log messages. To mitigate these issues, we recommend users employ trusted LLMs as backbones and filter out harmful code candidates carefully"[A76]
\\ 

&
& ``LLMs can be subject to bias [52, 53] and may generate unwanted discriminatory or offensive text. Moreover, it cannot be excluded that LLMs could be subject to adversarial attacks, leading to the generation of unwanted outputs as it has been shown for other recommender systems [89]."[A77]
\\ 
\end{longtblr}
\end{landscapeLongTblr}

\begin{longtblr}[
    caption={Table mapping potential risks themes to maintainability and evolvability quality attributes. Numbers in parentheses represent the number of primary studies within each theme associated with a given attribute.},
    label={tab:theme_attribute_risks}
]{
    colspec = {Q[l,m,6cm] Q[l,t,2cm] X[l,m]},
    vlines,
    hlines,
    hline{1-2}= {2pt,solid},
    rows = {font = \fontsize{8pt}{9pt}\selectfont}
}
\textbf{RQ2 Theme}  & \textbf{Quality Construct} & \textbf{Related Attributes} \\ 

% -- First High Order Theme
\SetCell[r=2]{} Methodological and Evaluation Challenges in LLM Research
& Maintainability
& Analyzability (5), Testability(4), Maintainability compliance (3), Stability (1), Changeability (1) \\

& Evolvability
& Extensibility (2), Changeability (2), Integrity (2), Testability(2), Analyzability (1) \\

% -- Second High Order Theme
\SetCell[r=2]{} LLM Output Quality, Accuracy, and Maintainability
& Maintainability
& Analyzability (7), Maintainability compliance (3), Testability(3), Changeability (3), Stability (3), Portability (2) \\

& Evolvability
& Changeability (8), Extensibility (5), Integrity (3), Analyzability (1) \\

% -- Third High Order Theme
\SetCell[r=2]{} Ethical, Security, and Trustworthiness Concerns in LLM Usage
& Maintainability
& Analyzability (3) \\

& Evolvability
& - \\

% -- Fourth High Order Theme
\SetCell[r=2]{} Practical Deployment and Contextual Adaptation Challenges
& Maintainability
& Analyzability (3), Stability (2), Maintainability compliance (2), Testability(1), Changeability (1)  \\

& Evolvability
& Changeability (8), Integrity (3), Extensibility (1) \\

% -- Fith High Order Theme
\SetCell[r=2]{} Impact of Training Data on AI Model Performance
& Maintainability
& Analyzability (1), Testability(1), Maintainability compliance (1) \\

& Evolvability
& Changeability (1) \\

\end{longtblr}

\subsection{RQ4 -- What weaknesses do LLMs exhibit with respect to software evolvability and maintainability?}

To address RQ4, we extracted quotes describing weaknesses of LLMs that frequently persist across approaches. In analyzing these quotations, we sought to identify more structural shortcomings of current LLM technologies. Table~\ref{tab:weaknesses} shows an overview of the generated themes, papers related, and some exemplary quotes categorized in those themes. A total of 110 quotations were synthesized into the following themes:

\begin{enumerate}
    \item Challenges in AI-Generated Code Quality and Correctness: 34 quotes
    \item Human-AI Interaction, Prompting and Evaluation in Software Development: 25 quotes
    \item AI's Conceptual Understanding and Generalizability Limitations: 24 quotes
    \item Methodological and Research Gaps in AI for Software Engineering: 18 quotes
    \item AI System Performance and Resource Management: 8 quotes
    \item Ethical and Human Oversight Considerations in AI-Assisted Development: 1 quote.
\end{enumerate}

The most prominent weakness concerns the \textit{AI-generated code quality and correctness}. A relevant number of studies report compilation errors, runtime failures, incorrect logic, and violations of expected behavior in generated solutions. These findings indicate that \textbf{LLMs often struggle to reliably preserve correctness and do not consistently produce high-quality outputs when modifying or creating software artifacts}. From a maintainability perspective, such defects may increase debugging effort and correction costs, while from an evolvability standpoint, they hinder safe and confident system evolution. 

The same theme includes several studies that highlight weaknesses related to design integrity and architectural coherence. LLM-generated code may introduce unnecessary complexity, redundant constructs, or poor separation of concerns, making systems harder to understand and evolve over time. These findings indicate that \textbf{LLMs often lack an understanding of system architecture, which is critical for long-term evolvability}.

There is also a substantial number of reported weaknesses related to \textit{human-AI interaction, prompting and evaluation in software development:}. Many studies indicate that \textbf{variations in task specification, context framing, or prompt wording can lead to dramatically different outputs}. This sensitivity may increase developers’ cognitive workload, as they must carefully craft prompts to obtain consistent and reliable results. Such dependence can make automation difficult and may undermine reliability, particularly in continuous integration settings or large-scale development workflows.

Another major weakness lies in the \textit{AI’s conceptual understanding and generalizability limitations}. A large number of quotations indicate that models frequently fail when tasks require deep reasoning about software abstractions, domain-specific constraints, or architectural intent. This shallow understanding limits their ability to generalize beyond familiar patterns observed during training. As a result, \textbf{LLM assistance may degrade in scenarios where evolvability depends on informed design decisions rather than surface-level code transformations}.

A large number of works also highlight \textit{methodological and research gaps in AI for software engineering}. Studies note the absence of standardized benchmarks, inconsistent baselines, and limited comparability across evaluations. In some cases, insufficient reporting of experimental settings, datasets, or model configurations further complicates the interpretation and replication of results. These methodological weaknesses not only \textbf{hinder cumulative scientific progress but also obscure the true limitations of LLMs, making it difficult to assess their long-term impact on maintainability and evolvability}.

\textit{AI system performance and resource management} is associated with a smaller set of quotations. These include scalability limitations when handling large codebases, constraints imposed by long contexts or token limits, and large-scale projects that require substantial computational resources or incur high financial costs. Such performance-related issues \textbf{may limit the applicability of LLMs to localized tasks and can introduce challenges when maintaining and evolving large software systems}.

The final theme, consisting of a single quotation, addresses \textit{ethical and human oversight considerations in AI-assisted development}. Although far less frequent than technical concerns, this observation emphasizes issues of ownership, responsibility, and the need for transparent human review of AI-generated artifacts. 

In summary, the weaknesses reported across the reviewed studies present a multifaceted picture of the current constraints facing LLMs in software evolution and maintenance. The most recurrent issues involve low-quality or incorrect code generation, unreliable bug fixing, and difficulties in preserving core design properties, often exacerbated by models’ dependence on precise and context-rich prompts. Limitations in conceptual understanding, domain reasoning, and generalizability further restrict applicability in specialized or complex scenarios. These technical challenges are compounded by methodological shortcomings in existing research, performance and scalability constraints, and emerging ethical concerns related to ownership and oversight. Taken together, these findings delineate the current boundaries of LLM capabilities and highlight key factors that may hinder their effective use in long-term software evolution.

Table~\ref{tab:theme_attribute_weaknesses} synthesizes how the weakness-related themes identified in RQ4 map onto the maintainability and evolvability attributes adopted in this study. The mapping shows similar results from the previous sections where they are predominantly associated with \emph{analyzability}, \emph{stability}, and \emph{maintainability compliance} within the maintainability construct, highlighting concerns related to inconsistent reasoning and limited domain understanding that complicate fault diagnosis and undermine trust in generated artifacts. From an evolvability perspective, the identified weaknesses mainly relate to \emph{integrity}, \emph{changeability}, and \emph{extensibility}, indicating that structural deficiencies in LLM behavior may erode architectural coherence and introduce changes that are difficult to sustain over time. Overall, the table clarifies that the weaknesses reported in RQ4 represent systemic challenges that directly threaten core maintainability and evolvability attributes, reinforcing the need for mitigation strategies discussed in the subsequent section.

\begin{landscapeLongTblr}
\begin{longtblr}[
    caption={Table showing all higher themes with their respective lower themes for the quotes on weaknesses, codes and a few examples of quotes that would fit their lower themes. The 'ID' column shows the id for the paper from where the quote was extracted. The full table of selected studies and their references is available in \cite{selectedStudies}.},
    label={tab:weaknesses}
]{
    colspec = {Q[l,m,2cm] Q[l,t,2cm] X[j,m]},
    vlines,
    hlines,
    hline{1-2}= {2pt,solid},
    rows = {font = \fontsize{8pt}{9pt}\selectfont}
}
\textbf{Higher-level Theme}  & \textbf{ID} & \textbf{Exemplary Quotes} 
\\
    
% -- First High Order Theme
\SetCell[r=3]{} Challenges in AI-Generated Code Quality and Correctness & \SetCell[r=3]{} A4, A6, A12, A17, A19, A23, A25, A29, A31, A34, A43, A44, A47, A62, A69, A70, A90, A91, A96, A98, A104 
& ``Auto-generated programs share common mistakes with human-written programs, and contain certain negative symptoms including: (1) names indicate wrong algorithms; (2) similar code blocks; (3) irrelevant helper functions."[A17]
\\

& 
& ``Existing pattern based and learning based APR are ineffective at fixing auto-generated code, challenges include: (1) limited search space; (2) unable to generate multi-edit patches; (3) lack of awareness of program dependencies."[A17]
\\ 

& 
& ``Although LLMs can take larger diffs as input, their performance of generating messages leaves much to be improved. UniXcoder tends to generate short messages, while ChatGPT can generate more detailed messages, which are very different from those written by developers."[A19]
\\

% -- Second High Order Theme
\SetCell[r=3]{} Human-AI Interaction, Prompting and Evaluation in Software Development
& \SetCell[r=3]{} A4, A12, A17, A31, A34, A40, A53, A58, A59, A61, A70, A73, A76, A85, A86, A100, A101, A102
& ``Effectiveness without guidance: When the vanilla LLM was only provided with the buggycode and commit message, its suggestions were less useful. Without guidance, the code snippets generated by the LLM lacked coherence and resulted in more failed test cases. Indeed, 26 out of 40 of these suggestions resulted in failed test cases. 14 of those failing cases were prevented when using CounterACT plans, showing that our generated plans can guide the LLM in providing more useful suggestions"[A58]
\\ 

&
& ``Another limitation we encountered was the lack of confidence scores for solutions within Copilot’s setup. Even though in our Copilot configuration, we set (ShowScore) to True, Copilot did not display the confidence intervals for each solution. Because of this constraint, we are unable to include this metric in our experimental results."[A12]
\\ 

&
& ``A full scale evaluation of the refactoring quality is challenging and could not be performed by us  at the time given that the scope of our study is the creation of a pipeline. This indicates that  a further configuration of our tested model ChatGPT is required, in order to successfully  automatically refactor data clumps"[A4]
\\ 

% -- Third High Order Theme
\SetCell[r=3]{} AI's Conceptual Understanding and Generalizability Limitations
& \SetCell[r=3]{} A11, A19, A23, A25, A40, A46, A51, A59, A77, A88, A92, A93, A95, A97, A99, A106
& ``We find that although GitHub Copilot and SATD exist at the bridge of natural language and programming language, there are characteristics of a TODO comments that do not align well with a generative prompt"[A11]
\\

& 
& ``Besides, the potential threat exists that the proposed testing and enhancing framework, CCTEST, may not adapt to other types of code completion systems."[A77]
\\

& 
& ``At the same time, recent research has pointed out perils of MLbased language translation [80], especially because the translation may not take into account that different programming languages may follow different programming paradigms (e.g., object oriented vs functional), and the result could just be “Java with a Python syntax” or something similar."[A77] \\

% -- Fourth High Order Theme
\SetCell[r=2]{} Methodological and Research Gaps in AI for Software Engineering
& \SetCell[r=2]{} A4, A26, A28, A29, A34, A36, A42, A48, A59, A61, A63, A87, A89, A103, A105
& ``This mechanism, however, is strongly dependent on the quality of the initial solution provided by the LLM. While related to the actual correctness of the solution, the quality needed to specialize the grammar is the ability for the LLM to provide a reasonable set of functions, libraries, and constants to be used by GE. This is generally an easier task for an LLM with respect to the production of correct code, thus making even smaller LLMs a good source of initial solutions for the GI-based step."[A36]
\\ 

& 
& ``the greedy algorithm employed in our approach to minimize the number of test oracles might not be the most optimal solution for minimizing test oracles while maximizing MS."[A34]
\\

% -- Fifth High Order Theme
\SetCell[r=2]{} AI System Performance and Resource Management
& \SetCell[r=2]{} A4, A34, A35, A81, A94
& ``The open-access API of Codex has a limit on the number of requests (20 per minute) and the number of tokens (40,000perminute). For this reason, our experiment needs to stop calling the API once in a while to not exceed the limit."[A34]
\\

& 
& ``During our experiments, we frequently reached the daily token limit when using ChatGPT"[A4]
\\

% -- Sixth High Order Theme
Ethical and Human Oversight Considerations in AI-Assisted Development 
& A77
& ``In general, there is a clear risk related to the ownership and understanding of code contributed via ChatGPT, especially when it is used to contribute complete features. Such a problem has been well-summarized in a comment of a PR we inspected [6]: “[. . . ] I want to make something clear about code suggestions done by ChatGPT: deferring to an AI bot is not the same as code ownership [. . . ] the idea that an author puts some code into a commit and sends it means they should have an intellectual understanding of it. PR authors should own the code they send – ownership in the sense of being able to advocate for the code."[A77]
\\ 
% -- End Table
\end{longtblr}
\end{landscapeLongTblr}

\begin{longtblr}[
    caption={Table mapping weaknesses themes to maintainability and evolvability quality attributes. Numbers in parentheses represent the number of primary studies within each theme associated with a given attribute. },
    label={tab:theme_attribute_weaknesses}
]{
    colspec = {Q[l,m,6cm] Q[l,t,2cm] X[l,m]},
    vlines,
    hlines,
    hline{1-2}= {2pt,solid},
    rows = {font = \fontsize{8pt}{9pt}\selectfont}
}
\textbf{RQ2 Theme}  & \textbf{Quality Construct} & \textbf{Related Attributes} \\ 

% -- First High Order Theme
\SetCell[r=2]{} Challenges in AI-Generated Code Quality and Correctness
& Maintainability
& Analyzability (11), Maintainability compliance (6), Testability(5), Portability (3), Stability (3), Changeability (2) \\

& Evolvability
& Changeability (8), Extensibility (4), Analyzability (3), Integrity (2), Testability(2), Portability (2) \\

% -- Second High Order Theme
\SetCell[r=2]{} AI's Conceptual Understanding and Generalizability Limitations
& Maintainability
& Analyzability (4), Maintainability compliance (3), Testability(2), Changeability (2), Stability (2), Portability (1) \\

& Evolvability
& Integrity (4), Changeability (3), Extensibility (2), Analyzability (2), Portability (2), Testability (1) \\

% -- Third High Order Theme
\SetCell[r=2]{} Human-AI Interaction, Prompting and Evaluation in Software Development
& Maintainability
& Analyzability (10), Maintainability compliance (6), Stability (6), Testability(4), Changeability (1)  \\

& Evolvability
& Integrity (5), Changeability (5),  Extensibility (3), Analyzability (2), Testability(1),  Portability (1)\\

% -- Fourth High Order Theme
\SetCell[r=2]{} AI System Performance and Resource Management
& Maintainability
& Testability(3), Analyzability (2), Stability (2), Changeability (1)  \\

& Evolvability
& Changeability (2), Extensibility (2), Integrity (1), Analyzability (1) \\

% -- Fith High Order Theme
\SetCell[r=2]{} Methodological and Research Gaps in AI for Software Engineering
& Maintainability
& Testability(7), Analyzability (6), Maintainability compliance (3), Stability (2), Changeability (2) \\

& Evolvability
& Integrity (5), Extensibility (4), Changeability (4), Analyzability (3) \\

% -- Sixth High Order Theme
\SetCell[r=2]{} Ethical and Human Oversight Considerations in AI-Assisted Development 
& Maintainability
& Analyzability (1) \\

& Evolvability
& - \\

\end{longtblr}

\subsection{RQ5 -- How can weaknesses of LLMs be mitigated to
better support software evolvability and maintainability?}

From the analyzed studies, we extracted 50 quotations describing solutions proposed to mitigate the weaknesses identified in RQ4. These solutions reflect efforts to present complementary strategies, encompassing not only technical fixes but also aspects related to human–model interaction, research practices, quality assurance, and operational considerations. Table~\ref{tab:solutions_all} presents the higher-level themes resulting from the synthesis, along with the associated paper identifiers and representative quotations for each theme. The five higher-level themes are as follows:

\begin{enumerate}
    \item Strategies for optimizing LLM performance and interaction: 22 quotes
    \item Ensuring methodological rigor and validity in AI for SE research: 12 quotes
    \item Post-generation quality assurance for AI-generated artifacts: 8 quotes
    \item Exploring future directions and novel applications of LLMs in software engineering: 3 quotes
    \item Addressing operational challenges in AI-driven software engineering: 3 quotes.
\end{enumerate}

Accounting for nearly half of all identified quotations, \textit{strategies for optimizing LLM performance and interaction} group observations indicating that many weaknesses attributed to LLMs, such as unstable outputs, misinterpretation of requirements, or hallucinations, are strongly influenced by how models are prompted and contextualized. The proposed solutions emphasize structured and systematic prompting practices, richer context provision, and the explicit incorporation of domain knowledge. Several studies also highlight the value of embedding procedural guidance or planning mechanisms into prompts, indicating that \textbf{careful interaction design can substantially improve reliability and output quality without modifying the underlying model}.

Rather than targeting the model itself, \textit{ensuring methodological rigor and validity in AI-for-SE research} addresses weaknesses arising from experimental design and evaluation practices. Authors recommend clearer documentation of model configurations and parameters, the use of higher-quality and more representative datasets, and validation strategies that account for variability across tasks and models. Collectively, these studies suggest that \textbf{some reported LLM weaknesses stem from inconsistent or fragile research practices, and that strengthening methodological foundations is essential for producing trustworthy and generalizable results}.

A third group of solutions highlights the importance of \textit{post-generation quality assurance for AI-generated artifacts}. These approaches acknowledge that even well-configured LLMs can produce incorrect or low-quality outputs, making downstream validation indispensable. Proposed mitigation strategies include automated checking mechanisms, static analysis, testing pipelines, and human-in-the-loop review processes. This theme directly responds to concerns raised in RQ4 regarding output correctness and maintainability, reinforcing the view that \textbf{LLM-assisted development should be embedded within robust quality assurance workflows}.

Less frequently discussed, but still relevant, are solutions related to \textit{exploring future directions and novel applications of LLMs in software engineering}. These quotes point toward research opportunities rather than immediate fixes, such as integrating LLMs more deeply with development environments or combining them with complementary AI techniques. Although limited in number, these proposals suggest longer-term pathways for overcoming current limitations.

Finally, a small set of studies addresses \textit{operational challenges in AI-driven software engineering}, including issues related to scalability, computational cost, and practical deployment constraints. While these concerns appear less prominently in the literature, they highlight that \textbf{mitigating LLM weaknesses also involves aligning technical solutions with real-world operational contexts}.

In summary, the solutions proposed depict an emerging yet coherent landscape that closely aligns with the weaknesses identified in RQ4. Most recommendations focus on optimizing interactions with LLMs, through improved prompting practices, better model tuning, and the incorporation of contextual or domain knowledge, suggesting that \textbf{many current limitations can be alleviated through more effective human–AI collaboration}. At the same time, solutions addressing methodological rigor, dataset quality, and post-generation validation underscore the \textbf{importance of robust processes to ensure reliability and reproducibility}. The relatively small number of forward-looking strategies suggests that long-term or structural improvements to LLM-based development workflows remain comparatively underexplored. Taken together, these findings delineate both the current maturity of mitigation strategies and the gaps that warrant further attention as LLMs become more deeply integrated into software engineering practice.

Table~\ref{tab:theme_attribute_solutions} synthesizes how the mitigation strategy themes identified in RQ5 map onto the maintainability and evolvability attributes considered in this study. The mapping indicates that the proposed remedies predominantly target \emph{analyzability} and \emph{maintainability compliance} within the maintainability construct, reflecting efforts to improve transparency, diagnosability, and adherence to development standards when integrating LLMs into software engineering workflows. Strategies such as human-in-the-loop validation, structured prompting, and hybrid pipelines are also associated with \emph{stability}, highlighting their role in reducing unintended side effects and increasing confidence in LLM-assisted modifications. From an evolvability perspective, the mitigation themes are mainly linked to \emph{integrity}, \emph{changeability}, and \emph{extensibility}, indicating that these strategies aim to preserve architectural coherence while enabling controlled and sustainable system evolution. Overall, the table clarifies that the remedies proposed in the literature directly operationalize maintainability and evolvability concerns, translating identified risks and weaknesses into concrete practices that support the long-term sustainability of LLM-assisted software systems.

\begin{landscapeLongTblr}
\begin{longtblr}[
    caption={Table showing all higher themes with their respective lower themes for the quotes on solutions for weaknesses, codes and a few examples of quotes that would fit their lower themes. The 'ID' column shows the id for the paper from where the quote was extracted. The full table of selected studies and their references is available in \cite{selectedStudies}.},
    label={tab:solutions_all}
]{
    colspec = {Q[l,m,2cm] Q[l,t,2cm] X[j,m]},
    vlines,
    hlines,
    hline{1-2}= {2pt,solid},
    rows = {font = \fontsize{8pt}{9pt}\selectfont}
}
\textbf{Higher-level Theme}  & \textbf{ID} & \textbf{Exemplary Quotes} 
\\ 
% -- First High Order Theme
\SetCell[r=3]{} Strategies for Optimizing LLM Performance and Interaction
& \SetCell[r=3]{} A4, A6, A16, A17, A25, A28, A29, A40, A46, A48, A59, A70, A86, A88, A90, A91, A93, A96, A97, A98, A102, A106
& ``Another approach is to incorporate repair-specific knowledge by using additional templates as demonstrated in Section V-D to reduce the amount of code LLM has to generate and arrive at the correct patch faster."[A16]
\\ 

& 
& ``More sophisticated prompt engineering, can be a useful method to improve the results, since it can be seen that the effect of the parameters is noticeable."[A4]
\\ 

& 
& ``Two potential directions for mitigating these issues were identified: improving the large language model, such as using GPT-4 instead of GPT-3.5, and improving the quality of reviews, such as providing more clear information."[A40]
\\ 

% -- Second High Order Theme
\SetCell[r=3]{} Ensuring Methodological Rigor and Validity in AI-in-SE Research
& \SetCell[r=3]{} A3, A4, A23, A29, A31, A61, A89, A92, A95, A99, A103, A105
& ``We intend to experiment with components identified using reverse engineering methods, aiming to assess the quality of summaries when the components express greater cohesion"[A3]
\\ 

& 
& ``A more thorough comparison between our proposed method and existing state-of-the-art techniques could enhance the significance of our results."[A4]
\\ 

& 
& ``To mitigate this problem, we perform human evaluation for verifying soundness of the revisions, albeit on a subset of our Python dataset, but ensuring full coverage in terms of the static checks"[A31] \\ 

% -- Third High Order Theme
\SetCell[r=3]{} Post-Generation Quality Assurance for AI-Generated Artifacts
& \SetCell[r=3]{} A12, A17, A23, A46, A87, A101, A104
& ``Our results highlight the importance for developers to continuously check the security of the code generated by such models through the implementation of rigorous security code reviews and with the use of a security analysis tool."[A12]
\\ 

&
& ``Instead of producing the correct program from scratch, future code generation can first produce an edit of the incorrect program, and further refine it via an iterative test-driven approach."[A17]
\\ 

&
& ``As our study shows that auto-generated programs with these symptoms are unlikely to lead to correct programs, future designers of language models can integrate a filter function into the language model to automatically eliminate programs with negative symptoms. Another alternative solution is to encode these symptoms into the ranking function to guide the language model in selecting better programs. Both of these directions indicate the potential of incorporating recent advancement of APR research in patch correctness assessment [27], [28] and patch prioritization [29], [30] to guide language models like Codex in generating better programs."[A17]
\\ 

% -- Fourth High Order Theme
\SetCell[r=3]{} Exploring Future Directions and Novel Applications of LLMs in Software Engineering
& \SetCell[r=3]{} A17, A29, A45
& ``Future language models designed for code generation should focus on summarizing useful information from problem description to reduce reliance on function names."[A17]
\\ 

& 
& ``leverage LLMs to create a high-quality benchmark dataset for commit message generation is worth further exploration."[A29]
\\ 

& 
& ``We need rigorous techniques to convert the fuzzy natural language processing capabilities of these models into precise, deterministic, and correct tools…"[A45]
\\ 

% -- Fifth High Order Theme
\SetCell[r=3]{} Addressing Operational Challenges in AI-Driven Software Engineering
& \SetCell[r=3]{} A4, A94, A100
& ``potential solution involves segmenting the project into smaller, more manageable parts."[A4]
\\

& 
& ``To increase efficiency, using smaller code search models trained with augmented data is recommended instead of directly using LLMs like ChatGPT at runtime."[A94]
\\ 

& 
& ``Development teams must weigh whether creating a new recipe […] is the most efficient use of resources.We are currently integrating LLMs in the process of writing OpenRewrite recipes."[A100]
\\
% -- End Table
\end{longtblr}
\end{landscapeLongTblr}

\begin{longtblr}[
    caption={Table mapping solution themes for identified weaknesses to maintainability and evolvability quality attributes. Numbers in parentheses represent the number of primary studies within each theme associated with a given attribute.},
    label={tab:theme_attribute_solutions}
]{
    colspec = {Q[l,m,6cm] Q[l,t,2cm] X[l,m]},
    vlines,
    hlines,
    hline{1-2}= {2pt,solid},
    rows = {font = \fontsize{8pt}{9pt}\selectfont}
}
\textbf{RQ2 Theme}  & \textbf{Quality Construct} & \textbf{Related Attributes} \\ 

% -- First High Order Theme
\SetCell[r=2]{} Strategies for Optimizing LLM Performance and Interaction
& Maintainability
& Analyzability (6), Maintainability compliance (6), Testability(6), Stability (4), Portability (2), Changeability (2) \\

& Evolvability
& Changeability (9), Integrity (6), Extensibility (4), Analyzability (2), Testability(1)\\

% -- Second High Order Theme
\SetCell[r=2]{} Ensuring Methodological Rigor and Validity in AI-in-SE Research
& Maintainability
& Analyzability (6), Maintainability compliance (3), Testability(3), Changeability (2), Stability (1), Portability (1) \\

& Evolvability
& Extensibility (6), Portability (3), Testability (2), Analyzability (1), Integrity (1), Changeability (1) \\

% -- Third High Order Theme
\SetCell[r=2]{} Post-Generation Quality Assurance for AI-Generated Artifacts
& Maintainability
& Analyzability (3), Stability (2), Testability(2), Maintainability compliance (1), Changeability (1)  \\

& Evolvability
& Changeability (4), Integrity (2), Analyzability (1), Portability (1)\\

% -- Fourth High Order Theme
\SetCell[r=2]{} Exploring Future Directions and Novel Applications of LLMs in Software Engineering
& Maintainability
& Analyzability (2)  \\

& Evolvability
& Extensibility (1) \\

% -- Fith High Order Theme
\SetCell[r=2]{} Addressing Operational Challenges in AI-Driven Software Engineering
& Maintainability
& Testability(1), Analyzability (1), Maintainability compliance (1), Stability (1), Changeability (1) \\

& Evolvability
& Changeability (2), Integrity (1), Extensibility (1), Testability (1) \\

\end{longtblr}

\section{Discussion}\label{sec5}

In this section, we interpret our findings through the lens of \textit{the Good, the Bad, the Ugly, and the Remedy}, a framing that captures both the opportunities and the limitations of using LLMs in the context of software maintainability and evolvability. This structure supports a balanced reflection on how LLMs can enhance maintenance- and evolution-oriented tasks, the risks and structural weaknesses they introduce to long-term software sustainability, and the mitigation strategies proposed to address these challenges.

\subsection{The Good: How LLMs Enable Sustainable Development}

The findings in Section~4.3 reveal clear and multifaceted benefits that LLMs can offer for software maintainability and evolvability. Across 140 positive-impact quotes synthesized into five higher-level themes, LLMs demonstrate substantial promise in enhancing developer productivity, improving code comprehension, supporting debugging activities, and strengthening software quality.

A central advantage lies in the broad applicability and adaptability of LLMs to diverse software engineering tasks. Studies document improvements in summarization, classification, code refinement, and even complex transformations, reflecting the versatility of current LLMs. Comparative evidence, where LLM-based approaches frequently outperform traditional baselines, underscores their effectiveness across various contexts.

Another important dimension of ``the Good'' involves the optimization and behavioral tuning of LLMs. Authors consistently report that small adjustments to prompts, parameters, or contextual information can significantly improve output quality, enabling LLMs to behave as adaptable assistants that respond well to guided input.

LLMs also show notable performance gains in code generation, analysis, and automated repair, with several studies highlighting robust patch generation, improved correctness, and enhanced fault localization. In addition, empirical evidence points to tangible benefits for developers: reduced cognitive load, decreased task completion time, and increased efficiency throughout the software development lifecycle.

Finally, the application of LLMs to testing and evaluation tasks further supports maintainable and evolvable systems by improving automated fault detection, test generation, and validation activities. Together, these findings illustrate that, when used strategically and with appropriate oversight, LLMs can strengthen core activities underpinning long-term software sustainability.

\subsection{The Bad: Risks and Failures that Threaten Maintainability and Evolvability}

The results in Section~4.4 reveal a series of practical risks that may threaten long-term maintainability and evolvability when LLMs are integrated into software development workflows. Synthesized across 48 quotes, these risks center on methodological shortcomings, inconsistent output quality, contextual failures, and issues originating in training data.

A major concern involves methodological and evaluation challenges. Many studies rely on datasets that may be outdated, unsafe, or even included in the training data of the evaluated models, raising threats of data leakage and compromised external validity. Furthermore, the use of automated metrics alone, without manual validation, may mask significant flaws in LLM-generated artifacts, increasing the danger of misinterpreting model performance.

Another critical set of risks relates to the accuracy, correctness, and maintainability of LLM-generated outputs. Evidence shows that LLMs frequently produce erroneous, misleading, or hallucinated content. As reported in one study (\cite{A77}), LLM suggestions were wrong in 36 out of 47 tested issues, illustrating the severity of the problem. These inaccuracies create new maintenance burdens, introduce instability into evolving systems, and risk accumulating new forms of technical debt.

Practical risks also emerge during deployment and contextual adaptation. LLMs often fail in the absence of precise, context-rich inputs, demonstrating brittle behavior when confronted with ambiguous requirements or domain-specific scenarios. Operational obstacles, such as computational cost, resource consumption, and scalability limitations, further complicate integration into real-world workflows.

Finally, the impact of training data introduces risks related to bias propagation, erroneous examples, or low-quality fine-tuning data. These issues can degrade the maintainability and evolvability of downstream systems by embedding flawed assumptions or unstable behavior into the generated artifacts.

\subsection{The Ugly: Structural Weaknesses Inherent to LLMs}

Section~4.5 focuses on the structural limitations of LLMs, which are not readily addressed through improved prompting, parameter tuning, or enhanced training data. The strongest concentration of evidence lies in the quality and correctness of AI-generated code, with 34 quotes reporting defects such as compilation failures, runtime errors, incorrect outputs, poor design integrity, and difficulty preserving cohesion or coupling. These issues are not occasional anomalies but systematic manifestations of how LLMs reason about code based on statistical associations rather than grounded semantic understanding.

A second structural limitation is the models' shallow conceptual understanding and weak generalizability. Across 24 quotes, studies note that LLMs struggle with deep programming concepts, domain-specific reasoning, or tasks requiring architectural insight. Their behavior often breaks down when confronted with unfamiliar frameworks, novel contexts, or tasks requiring reasoning beyond surface-level patterns.

Equally concerning are weaknesses in human--AI interaction and prompting, with 25 quotes documenting the models' extreme sensitivity to task specification. LLMs frequently produce inconsistent or misleading outputs when prompts are even slightly ambiguous. This dependence on precise instructions represents a structural constraint that places a high burden on developers and complicates integration into automated workflows.

Additional structural constraints arise from performance, scalability, and resource utilization, particularly when processing large inputs, long contexts, or tasks approaching token limits. These issues restrict the feasibility of applying LLMs to large-scale, real-world software systems.

Finally, a smaller yet important subset of studies points to ethical and oversight challenges, including concerns about authorship, responsibility, and the opacity of LLM-generated artifacts. Although less frequently discussed, these limitations reflect deeper socio-technical risks inherent to black-box generative systems.

Taken together, these findings portray a technology that is powerful but still fundamentally limited, well suited for assisting with certain maintenance tasks, yet prone to unpredictable behavior when pushed beyond its intrinsic architectural boundaries.

\subsection{The Remedy: Mitigating LLM Weaknesses}
\label{sec:remedy}

While \emph{the Good}, \emph{the Bad}, and \emph{the Ugly} characterize the opportunities, risks, and structural weaknesses associated with LLM use in software engineering, \emph{the Remedy} focuses on how the literature proposes to respond to some of these challenges. Rather than introducing fundamentally new capabilities, the mitigation strategies identified in RQ5 largely aim to constrain, compensate for, or contextualize existing LLM behavior in ways that better align with maintainability and evolvability goals.

A prominent observation emerging from this synthesis is that most remedies operate at the interaction, process, or workflow level, rather than at the level of core model architecture. Strategies such as structured prompting, richer contextualization, and the explicit incorporation of domain knowledge are frequently proposed as practical means to reduce output instability, misinterpretation of requirements, and hallucinations. These approaches suggest that many weaknesses identified in \emph{the Ugly} can be partially alleviated through more deliberate human--AI collaboration, reinforcing the role of developers as active supervisors rather than passive consumers of LLM outputs.

Beyond interaction-level interventions, a second group of remedies emphasizes methodological rigor and research validity. Improved reporting practices, standardized benchmarks, stronger baselines, and more representative datasets are repeatedly highlighted as prerequisites for producing reliable and generalizable evidence. This aligns closely with the methodological risks discussed earlier, indicating that some perceived weaknesses of LLMs may be artifacts of fragile evaluation practices rather than inherent model limitations. Strengthening research foundations is therefore positioned as a necessary condition for accurately assessing the long-term implications of LLM adoption for maintainability and evolvability.

A third set of remedies centers on post-generation quality assurance and human oversight. Automated testing, static analysis, validation pipelines, and human-in-the-loop review mechanisms are proposed as safeguards to detect and correct defects introduced by LLM-generated artifacts. These approaches directly address concerns raised in \emph{the Ugly} regarding correctness, architectural erosion, and the accumulation of technical debt. Importantly, they reflect an implicit acknowledgment that LLM outputs cannot be assumed to be reliable by default and must be embedded within robust assurance processes to support safe software evolution.

Despite these contributions, the synthesis also reveals notable gaps. Most remedies are reactive and localized, focusing on compensating for observed weaknesses rather than rethinking how LLMs are integrated into long-term software evolution workflows. Fewer studies explore structural or organizational interventions, such as evolving development processes, redefining accountability structures, or systematically aligning LLM use with architectural governance. As a result, while current remedies offer practical guidance for mitigating immediate risks, they provide limited insight into how LLM-assisted development can be sustainably scaled over time.

Taken together, \emph{the Remedy} highlights that mitigation strategies are advancing in parallel with the identification of risks and weaknesses, but remain uneven in scope and maturity. Existing approaches demonstrate that many challenges can be managed through careful interaction design, rigorous evaluation practices, and strong quality assurance. However, they also underscore the need for deeper, forward-looking research on how LLMs can be embedded into software engineering practices in ways that genuinely support long-term maintainability and evolvability, rather than merely offsetting their current limitations.

\subsection{Analysis Across the Four Dimensions}

Synthesizing insights from \textit{the Good, the Bad, the Ugly, and} \textit{the Remedy} reveals a landscape marked by both opportunity and tension. On one hand, Section~4.3 demonstrates that LLMs provide meaningful short-term benefits for maintainability and evolvability: they improve analyzability, support debugging and automated repair, assist in documentation and summarization, and reduce the cognitive effort required for common development tasks. These advantages suggest that LLMs can meaningfully strengthen the everyday practices that shape software quality.

However, the risks identified in Section~4.4 complicate this optimistic view. Many of these risks directly contradict the reported benefits. LLMs accelerate development, but the speed and convenience they introduce may conceal errors, inconsistencies, or hallucinations that erode long-term software sustainability. Improvements in code comprehension coexist with risks of incorrect or misleading explanations. Gains in developer productivity are counterbalanced by concerns that increased reliance on LLMs may accumulate new forms of technical debt, especially when generated
artifacts are integrated into evolving systems without sufficient verification.

These contradictions reflect an inherent tension: LLMs generate value primarily by making certain tasks easier and faster, yet they introduce new failure modes that are
harder to detect and may become costlier to correct over time. As such, \textbf{short-term maintainability benefits do not necessarily translate into long-term evolvability}. The \textbf{developer's role shifts from producing artifacts to validating them}, with the cost of verification increasing as system complexity grows.

The structural weaknesses described in Section~4.5 amplify this tension. \textit{The Ugly} dimension captures limitations that are not merely the result of immature tooling or evaluation methods, instead, they are fundamental to the current design of LLM architectures. A central concern is the lack of reliable semantic grounding in AI-generated code, which leads to recurring patterns of compilation failures, incorrect
logic, design inconsistency, and violations of architectural intent. These issues arise even when prompts are well crafted and contextual information is carefully provided.

In addition, LLMs exhibit shallow conceptual understanding, limited domain reasoning, and fragile generalizability. Their behavior can vary dramatically across tasks, frameworks, or datasets, undermining their usefulness in evolving systems that require stable abstractions and consistent architectural behavior. Weaknesses in human--AI interaction, particularly the extreme sensitivity to prompt formulation and task
specification, further complicate long-term maintenance and make automated usage difficult. These problems cannot be entirely mitigated by improved prompting or fine-tuning alone.

Performance and scalability issues compound the challenge, especially when LLMs operate near token limits or must process large inputs typical of industrial systems. The computational and financial costs associated with such workloads limit the
feasibility of large-scale integration in continuous evolution pipelines. Finally, although less frequently discussed, ethical and oversight concerns highlight sociotechnical risks
related to authorship, accountability, and trust in AI-generated artifacts.

The introduction of the Remedy dimension reframes this landscape by focusing on how these tensions can be actively managed. Rather than eliminating risks or structural weaknesses, the remedies identified in Section 4.6 emphasize compensatory strategies, such as structured interaction practices, methodological rigor, and post-generation quality assurance, that aim to contain failure modes and better align LLM use with maintainability and evolvability objectives.

Taken together, these insights reveal that the interplay between benefits, risks, structural weaknesses, and mitigation strategies determines whether LLMs ultimately support or hinder the sustainability of software systems. \textit{The Good} shows that LLMs can augment developers and reduce short-term effort. \textit{The Bad} demonstrates that these gains may be nullified or reversed when outputs propagate errors, inconsistencies, or hidden liabilities. \textit{The Ugly} exposes fundamental limitations that challenge the scalability, reliability, and
interpretability of LLM-enabled software evolution. \textit{The Remedy} highlights that sustainable adoption depends on embedding LLMs within robust processes that explicitly acknowledge and manage these limitations.

Balancing these forces requires reframing LLMs as fallible collaborators rather than automated solutions. Effective integration demands layers of validation, rigorous testing, human oversight, and prompt design strategies that incorporate contextual awareness and domain grounding. Such integration further depends on institutional and methodological commitments that ensure LLM-assisted workflows remain auditable, reproducible, and adaptable over time. It also requires recognizing the contexts in which LLMs excel, well-scoped, repetitive, or assistive tasks, versus those in which their
limitations become pronounced, such as complex architectural reasoning or safety-critical development. Ultimately, these tensions underscore a fundamental insight:
\textbf{LLMs can strengthen short-term maintainability, but without safeguards, they may undermine long-term evolvability}. \textit{The Remedy} therefore plays a central role in transforming LLMs from opportunistic productivity tools into dependable contributors to long-term software sustainability. Understanding and balancing these trade-offs is essential as LLMs become increasingly integrated into modern software engineering practice. 

\section{Threats to Validity}\label{sec6}

As with any systematic literature review, this study is subject to several threats to validity. We describe the main threats and the mitigation strategies adopted throughout the research process. 

\textbf{Construct Validity.} A primary concern relates to the interpretation of key concepts such as maintainability, evolvability, and LLM-related impacts. Variability in researchers’ understanding of these constructs may introduce inconsistencies in how quotes were identified or how attributes were assigned. Although we grounded our classification in established definitions (ISO/IEC 9126 and Breivold et al.), several attributes, such as analyzability, testability, and changeability, are conceptually shared across both constructs, and their practical meaning depends heavily on contextual interpretation. As a result, the same empirical observation could plausibly be framed as a short-term maintenance concern or as a long-term evolvability issue, depending on the analytical lens adopted. While we mitigated this subjectivity through calibration rounds, independent extraction, adjudication, and consensus-building discussions, some degree of interpretive judgment remains unavoidable. This reflexive limitation should be considered when interpreting the reported distributions of attributes and the boundaries drawn between maintainability- and evolvability-related evidence

\textbf{Internal Validity.} Human interpretation bias is inherent in qualitative research. Although each researcher independently extracted data from their assigned subset, differences in judgment may still arise. We addressed this threat through a cross-review process in which a designated adjudicator verified the consistency of all extracted data, and disagreements were resolved through discussions. In the thematic synthesis, the exploratory use of an LLM introduced potential model-driven biases. To reduce this threat, the LLM was used only for mechanical tasks (e.g., initial code generation), and all outputs were validated by the research team using a human-in-the-loop approach.

An additional methodological limitation concerns our choice of LLM for supporting the thematic synthesis process. We employed the gemini-2.5-flash model due primarily to its free availability, stable API access, and practical suitability for large-scale iterative coding under token and cost constraints. These pragmatic considerations were essential to enable reproducibility and to sustain repeated human-in-the-loop interactions across hundreds of qualitative excerpts. However, this choice introduces potential limitations. The thematic structure proposed by the LLM may reflect model-specific inductive biases and representational limitations rather than purely data-driven regularities.

We mitigated these risks through a hybrid human-in-the-loop process in which all generated codes were inspected, refined, merged, or rejected by the research team, and through calibration rounds and adjudication procedures to promote consistency and theoretical alignment. Nevertheless, the use of a single LLM model, chosen for pragmatic rather than methodological optimality, constitutes a reflexive limitation. Different models or configurations could plausibly have produced alternative intermediate codings or theme groupings, even if the high-level conclusions remained broadly similar. This dependency underscores that the LLM-assisted synthesis pipeline should be interpreted as a support mechanism rather than an objective or model-agnostic analytic process.

\textbf{External Validity.} The generalizability of our findings may be limited by the rapidly evolving landscape of LLM research. Several primary studies relied on specific LLM versions, experimental datasets, or prototype tools, which may hinder broader extrapolation. 

Another key threat to the external validity of this study stems from the maturity and nature of the available empirical evidence on LLM-assisted software engineering. Most primary studies in our corpus are short-term, task-oriented, and laboratory-based evaluations conducted on small or medium-sized open-source projects, synthetic benchmarks, or narrowly scoped datasets. Consequently, the evidence provides only limited insight into how LLM usage affects maintainability and evolvability over extended time horizons. Longitudinal effects, such as the accumulation of technical debt, architectural erosion, degradation of design intent, or shifts in maintenance effort, remain largely unobserved and empirically unmeasured. This constrains the extent to which current findings can substantiate claims about long-term software sustainability.

In addition, industry-grade evidence is scarce. Few studies examine LLM adoption within large-scale, regulated, or safety-critical industrial settings, where governance, compliance, auditability, security, and organizational dynamics play a decisive role. Most evaluations are conducted under controlled or semi-controlled conditions, often with carefully curated prompts and highly motivated users, which may not reflect realistic organizational usage patterns. As a result, reported productivity gains and quality improvements may not generalize to heterogeneous teams, legacy systems, or long-lived software ecosystems. These limitations highlight a critical research gap and motivate future longitudinal and industrial field studies to establish whether and under what conditions LLM-assisted software engineering genuinely supports sustainable software evolution, rather than merely shifting maintenance burdens into the future.

Nevertheless, the final corpus spans a variety of software development activities, LLM families, and empirical settings, supporting reasonable analytical diversity.

\textbf{Reliability.} Reproducibility may be affected by incomplete reporting in some primary studies, particularly regarding LLM model versions, configurations, or evaluation procedures. To enhance dependability, all extracted data and verbatim quotations were stored in a shared spreadsheet to ensure full traceability. The LLM-assisted thematic synthesis pipeline was configured with documented prompts, chunking strategies, and parameter settings. However, as LLMs evolve, future versions may produce different outputs, which represents an emerging challenge to reliability in empirical software engineering research.

Overall, we implemented multiple strategies, including calibration sessions, independent extraction, adjudication, explicit coding guidelines, and human validation of LLM-assisted synthesis, to mitigate threats across all validity dimensions. Nonetheless, due to the interpretive nature of qualitative research and the pace of LLM innovation, residual risks cannot be entirely eliminated.

\section{Conclusion}\label{sec7}

This study provides the first comprehensive synthesis of how LLMs influence the
maintainability and evolvability of software systems. By systematically reviewing the
existing empirical evidence and conducting a thematic analysis of 87 primary studies,
we mapped the quality attributes addressed in LLM-assisted software engineering and
synthesized the reported positive impacts, risks, weaknesses, and mitigation strategies
that influence the sustainability of LLM-enabled development workflows.

Our findings show that LLMs can meaningfully support long-term software
sustainability by enhancing analyzability, testability, code comprehension, repair
effectiveness, and developer productivity. Their broad applicability across software
life-cycle activities, combined with strong zero-shot and few-shot capabilities, 
positions LLMs as versatile assistants capable of reducing effort and cognitive load.
However, the evidence simultaneously reveals substantial risks. Incorrect or low-quality
code generation, hallucinations, prompt sensitivity, limited domain reasoning, and
inconsistent performance across tasks and datasets threaten maintainability and
evolvability. Methodological issues such as dataset contamination, weak baselines,
operational cost, and emerging ethical concerns further underscore the need for caution
when integrating LLMs into real-world engineering environments.

The weaknesses identified across studies highlight the need for systematic safeguards
when deploying LLM-based solutions. Proposed mitigation strategies center on improved
prompting practices, domain-aware model conditioning, iterative and test-driven
verification workflows, post-generation quality assurance mechanisms, and stronger
methodological standards for AI-for-SE experimentation. These findings collectively
suggest that hybrid strategies, where LLM outputs are validated, augmented, or
constrained by established software engineering techniques, are essential for ensuring
reliability and long-term sustainability.

Overall, the landscape revealed by this review is one of rapid change: LLMs introduce
important opportunities while also posing new sources of technical debt, dependency,
and risk. Advancing the state of the art will require more robust empirical methodologies,
improved evaluation protocols, LLM-aware quality metrics, and socio-technical processes
that govern responsible adoption. As LLM technologies continue to evolve, establishing
rigorous, evidence-based guidance becomes increasingly critical for researchers,
practitioners, and tool builders seeking to leverage these capabilities effectively and
ethically.

\subsection{Implications for Research and Practice}

The findings carry important implications for both the software engineering research
community and practitioners adopting LLM-based tools in real-world settings. For
researchers, this review exposes the scarcity of rigorous, transparent, and
context-sensitive empirical studies. Much of the current evidence is task-specific or
evaluated under idealized conditions that do not reflect industrial complexity. Future
work should emphasize longitudinal and cross-domain studies, replications, and designs
that explicitly measure evolvability and maintainability outcomes. Persistent
methodological concerns, such as dataset contamination, inconsistent evaluation
protocols, and unclear reporting, highlight the need for standardized, LLM-aware
research frameworks, stronger benchmarks, and better theoretical grounding that links
LLM capabilities to established quality models and socio-technical factors.

For practitioners, the evidence indicates that LLMs offer clear advantages in code
comprehension, testing support, and routine development acceleration, but these
benefits come with nontrivial risks. Organizations should position LLMs as assistive
tools rather than autonomous agents, implementing systematic validation mechanisms
to manage hallucinations, unstable outputs, and hidden defects. Effective adoption
requires human-in-the-loop workflows, structured prompting, automated quality checks,
and traceability mechanisms. Developer training in prompt engineering, evaluation
literacy, and responsible oversight is essential. Finally, as LLM-generated artifacts may
introduce new forms of technical debt, teams should explicitly incorporate LLM-specific
quality reviews and evolution planning into their development processes.

Together, these implications point toward a future in which LLMs play an increasingly
central role in software development while reinforcing the need for carefully designed
socio-technical practices and robust empirical foundations.

\bibliography{sn-bibliography}
\bibliographystyle{plain}

\end{document}